\documentclass[aps,amssymb,floatfix,prd,preprintnumbers]{revtex4-2}
\usepackage{epstopdf}
\usepackage{capt-of}
\usepackage{graphicx}  
\usepackage{dcolumn}   
\usepackage{bm}
\usepackage{amsmath}
\usepackage{multirow}
\usepackage[font=scriptsize]{caption}
\usepackage[colorlinks]{hyperref}
\usepackage{orcidlink}

\begin{document}

\newcommand{\jls}[1]{{\bf\textcolor{blue}{[Jackson: #1]}}}
\newcommand{\bmm}[1]{{\bf\textcolor{ForestGreen}{[Bivudutta: #1]}}}
\newcommand{\udt}[3]{#1^{#2}_{\phantom{#2}#3}}
\newcommand{\udut}[4]{#1^{#2\phantom{#3}#4}_{\phantom{#2}#3\phantom{#4}}}
\newcommand{\ududt}[5]{#1^{#2\phantom{#3}#4}_{\phantom{#2}#3\phantom{#4}#5}}
\newcommand{\dut}[3]{#1_{#2}^{\phantom{#2}#3}}
\newcommand{\dudt}[4]{#1_{#2\phantom{#3}#4}^{\phantom{#2}#3}}

\title{\bf Accelerating Cosmological Models in $f(T,B)$ Gravitational Theory}

\author{S. A. Kadam\orcidlink{0000-0002-2799-7870}}
\email{k.siddheshwar47@gmail.com}
\affiliation{Department of Mathematics, Birla Institute of Technology and Science-Pilani,\\ Hyderabad Campus, Hyderabad-500078, India.}
\author{Jackson Levi Said\orcidlink{0000-0002-7835-4365}}
\email{jackson.said@um.edu.mt}
\affiliation{Institute of Space Sciences and Astronomy, University of Malta, Malta, MSD 2080}
\affiliation{Department of Physics, University of Malta, Malta}

\author{B. Mishra\orcidlink{0000-0001-5527-3565} }
\email{bivu@hyderabad.bits-pilani.ac.in }
\affiliation{Department of Mathematics, Birla Institute of Technology and Science-Pilani, \\ Hyderabad Campus, Hyderabad-500078, India.}

\begin{abstract}
In this paper, we have explored the field equations of $f(T,B)$ gravity as an extension of teleparallel gravity in an isotropic and homogeneous space time. In the basic formalism developed, the dynamical parameters are derived by incorporating the power law and exponential scale factor function. The models are showing accelerating behaviour and approaches to $\Lambda$CDM at late time. The present value of the equation of state parameter for both the cases are obtained to be in accordance with the range provided by cosmological observations. The geometrical parameters and the scalar field reconstruction are performed to assess the viability of a late time accelerating Universe. Further the stability of both the models are presented. It has been observed that both the models are parameters dependent. Since most of the geometrically modified theories of gravity are favouring the violation of strong energy condition, we have derived the energy conditions both for the power law and exponential model. In both the models, the violation of strong energy condition established. 
\end{abstract}

\maketitle
{\bf{Keywords:}} $f(T,B)$ gravity, Accelerating model, Stability analysis, Energy conditions.
\section{Introduction}

The late time accelerating Universe \cite{Perlmutter99, Riess98} has prompted an enormous amount of research in the literature, which has largely been directed at understanding the properties of dark energy (DE). Hence, the effort to modify General Relativity (GR) has become necessary and at the first instance, the modification has been done in the geometrical part of Einstein-Hilbert action. One modification is the introduction of the Ricci curvature scalar $\mathring{R}$ in Einstein-Hilbert action leads to the $f(\mathring{R})$ gravity. Another one is the  extensions of teleparallel gravity leads to $f(T)$ gravity \cite{Linder10}, where $T$ is the torsion scalar. The approach in $f(T)$ gravity is to use the teleparallel connection \cite{Weitzenbock23, Bahamonde21,Bajardi21, Basilakos13}, which has the torsion in stead of curvature which is embodied in the Levi-Civita connection. Ref.~\cite{Myrzakulov11} suggests reconstruction methods to study the late time acceleration cosmological models in Friedmann–Lema\^{i}tre–Robertson–Walker (FLRW) space-time. In Ref.~\cite{Capozziello11} cosmographic test were performed, which is a model independent process, in $f(T)$ gravity. Another example of this non-parametric approach is Gaussian processes which have been performed in teleparallel gravity in Refs.~\cite{Briffa20, Said21, Cai20}. \\

On the other hand, Ref.~\citep{Cai16} reviews the cosmological solutions based on $f(T)$ gravity and discussed various ideas of its implications. The result obtained in Ref.~\cite{Awad17} in the extended teleparallel gravity have indicated the possibility of contracting phase prior to the expanding phase of the universe, while Ref.~\cite{Channuie18} derives the Noether equations of the non-local theory in flat FLRW universe and have analysed the dynamics of the field in a non-local  gravity that has been admitted by the Noether symmetry. In Ref.~\cite{Mirza19} it was shown that mimetic theory in $f(T)$ gravity can be formulated with the Lagrange multiplier method  without using any auxiliary metric. In $f(T)$ gravity, Ref.~\cite{Golovnev20} explains the Bianchi identities and have shown its compatibility on the corresponding equations. Several other ideas in the direction of teleparallel gravity can be seen in Ref. \cite{Ferraro08, Li11, Ferraro11, Nashed13, Izumi13, Paliathanasis14, Bejarano17, Krssak16, Bose20, Jimenez21, Duchaniya22,Capozziello15,Capozziello16,Kadam22EPJC}.\\

One of the first generalizations of GR came in the form of Brans-Dicke theory, and subsequently the $f(\mathring{R})$ gravity \cite{Buchdahl70}, which is a fourth order theory. Another intriguing idea was to change the connection by which gravity is expressed, namely from the curvature-based Levi-Civita connection to the torsion-based teleparallel connection. So, an idea has come up to consider $f(T,B)$ gravity theory, where $T$ and $B$ are two contributing scalars, known as the torsion scalar and the boundary term. The torsion scalar $T$ and boundary term $B$ exhibit the second-order and fourth-order derivative contributions respectively. Hence the $f(T,B)$ gravity is the generalization of $f(\mathring{R})$ and $f(T)$ gravity theory. Along a similar vein, Ref.~\cite{Bahamonde15,Bahamonde17} generalizes the $f(T)$ gravity with $f(T,B)$ gravity by incorporating the boundary term $B$. The boundary term is related to the divergence of the torsion tensor. For the choice of the functional $f(-T+B)$, the $f(T,B)$ gravity reduces to $f(\mathring{R})$ gravity. In recent years several cosmological aspects have been studied with $f(T,B)$ gravity. By exploiting the breaking of conformal symmetry, the study of  primordial magnetic fields with a non-adiabatic behavior has been studied in $f(T,B)$ gravity \cite{Capozziello22}. The study of the polarization and helicity of the gravitational waves made in Ref.\cite{Capozziello20} and show that $f(T,B)$ gravity shows three polarization mode. Using cosmological reconstruction techniques Ref.~\cite{Bahamonde18} shows that $f(T,B)$ gravity can mimic de sitter universe, power law and $\Lambda$CDM models. Using this approach Ref.~\cite{Caruana20} has presented the bouncing solution in this extended teleparallel gravity and explored the singularity and little rip cosmology, while Ref.~\cite{Franco20} performed the stability analysis on the cosmological models that has been modelled in confrontation with the observational data for the accelerating universe. On the observational side of this class of models, Ref.~\cite{Rivera20} has shown four cosmological models that have shown promise in meeting the late time cosmic acceleration measurements which can produce quintessence behaviour and experience transition along the phantom-divide line.\\

Ref.~\cite{Pourbagher20} has shown the thermodynamic effect of $f(T,B)$ gravity with the matter field in the form of viscous fluid. With the choice of jerk parameter, Ref.~\cite{Zubair20} studies the cosmological significance of $f(T,B)$ gravity, while Ref.~\cite{Moreira21} explores the five dimensional $f(T,B)$ gravity and revealed that the splitting brane process satisfy the strong and weak energy condition for the representative values of the model parameters. In Ref.~\cite{Sahlu21} it was shown that $f(T,B)$ gravity can explain the accelerating universe using power law cosmology. Finally, in Ref.~\cite{Bhattacharjee21,Kadam22} the energy conditions were constrained by choosing appropriate parametric value and shown the violation of strong energy conditions.\\
 
The paper is organised as follow: in Sec.~\ref{sec:background}, the basic field equations of $f(T,B)$ gravity in FLRW space-time has been derived, the effective pressure and effective energy density are expressed with respect to the Hubble parameter. In Sec.~\ref{sec:power_law_cosmo}, the presumed power law cosmology has been presented and the exponential scale factor model in Sec.~\ref{sec:exponential_Scale_factor}. The dynamical parameters, energy conditions, scalar field reconstruction and stability of both the models are discussed in their respective section. Finally the conclusion of both the models are given  Sec.~\ref{sec:conclusion}.

\section{\texorpdfstring{$f(T,B)$}{} Gravity Field Equations}\label{sec:background}

Modifications and extensions using the Ricci scalar has been instrumental in modifying gravity in many scenarios \cite{Clifton12}. In teleparallel gravity, the Levi-Civita connection used in GR is replaced by the teleparallel connection $\Gamma_{\mu \nu}^{\sigma}$. The Levi-Civita connection has non-zero curvature but zero torsion whereas, the teleparallel connection has non-zero torsion but zero curvature, and satisfies the metricity condition \cite{Cai16}. To this end, a connection with zero curvature resulted in vanishing the Riemann tensor. This is because the teleparallel gravity requires the bottom-up construction of different tensorial quantities to produce theories of gravity.  It is worthwhile to mention here that the dynamical objects in teleparallel gravity made up of tetrad $e_{\mu}^a$ in place of the metric $g_{\mu \nu}$ that was used in GR and some other modified theories of gravity. Now, the tetrads transform between manifold and Minkowski space indices with the following expression,

\begin{eqnarray}\label{eq:1}
g_{\mu\nu}&=&\eta_{ab}\udt{e}{a}{\mu}\udt{e}{b}{\nu}, \nonumber \\
\eta_{ab}&=& \dut{e}{a}{\mu}\dut{e}{b}{\nu} g_{\mu \nu},
\end{eqnarray}
where $\eta_{ab}$ represents the Minkowski metric, and $g_{\mu\nu}$ is the metric for the general manifold. The tetrad fields are orthonormal vector at each point of the manifold and for consistency it obeys the following orthogonality conditions,
\begin{eqnarray}
    \udt{e}{a}{\mu}\dut{e}{b}{\mu}&=&\delta^{a}_{b}\,, \nonumber\\
    \udt{e}{a}{\mu}\dut{e}{a}{\nu}&=&\delta^{\nu}_{\mu}\,.\label{eq:2} 
\end{eqnarray}

Now, the teleparallel connection $\Gamma_{\mu \nu}^{\sigma}$ connection can be expressed with respect to the tetrads and spin connection $\omega_{b\mu}^{a}$ as \cite{Weitzenbock23, Bahamonde21, Cai16} through

\begin{equation}
    \Gamma^{\sigma}_{\nu\mu} := \dut{e}{a}{\sigma}\left(\partial_{\mu}\udt{e}{a}{\nu} + \udt{\omega}{a}{b\mu}\udt{e}{b}{\nu}\right)\,,
\end{equation}
where the spin connection represents the degrees of freedom associated with the local Lorentz transformation invariance of the theory, and which are zero in the so-called Weitzenb\"{o}ck gauge of the connection. We have already mentioned that for the limit $f(T,B)=f(-T+B)=f(\mathring{R})$, meaning that the theory reduces to $f(\mathring{R})$ gravity, where $\mathring{R}$ is the regular Levi-Civita Ricci scalar. More generally, the action for $f(T,B)$ gravity can be given as,
\begin{equation}
    S_{f(T,B)}=\int d^4x e\mathcal{L}_m+\frac{1}{2\kappa^2}\int d^4x  e f(T,B)\,,\label{eq:4}
\end{equation}
where $e$ is the determinant of the tetrad, and $\kappa^2= 8\pi G$. Subsequently, varying the action in Eq.~\eqref{eq:4} with respect to the tetrad fields, the $f(T,B)$ gravity field equations can be derived as \cite{Bahamonde21, Bahamonde17}
\begin{eqnarray}
& & e_a{}^{\mu} \square f_B -  e_a {}^{\nu} \nabla ^{\mu} \nabla_{\nu} f_B +
	\frac{1}{2} B f_B e_a{}^{\mu} - \left(\partial _{\nu}f_B + \partial
	_{\nu}f_{T} \right)S_a{}^{\mu\nu}  \nonumber \\
& & -\frac{1}{e} f_{T}\partial _{\nu} (eS_a{}^{\mu\nu})
	+ f_{T} T^{B}{}_{\nu a}S_{b}{}^{\nu\mu}- f_T \omega ^b{}_{a\nu}
	S_b{}^{\nu\mu} -\frac{1}{2}  f E_{a}{}^{\mu} =  \kappa ^2  \Theta _a{}^{\mu} \,,\label{eq:5}
\end{eqnarray}
where $f_T$ and $f_B$ respectively partial derivative with respect to $T$ and $B$. Also, $\Theta_{a}^{~~\mu}$ and $\nabla_{\nu}$ are respectively denote the energy momentum tensor and Levi-Civita covariant derivative with respect to the Levi-Civita connection. now considering a tetrad for the flat FLRW metric, taken as
\begin{equation}\label{eq:6}
    e_{\mu}^{a}=(1,a(t),a(t),a(t))\,,
\end{equation}
which satisfies the Weitzenb\"{o}ck gauge for $f(T,B)$ gravity. We wish to study the cosmological aspects of $f(T,B)$ gravity in an isotropic and homogeneous background, which through Eq.~\eqref{eq:1} reproduces,
\begin{equation}\label{eq:7}
    ds^2=-dt^2+a^2(t)(dx^2+dy^2+dz^2)\,.
\end{equation}

We consider that a universe filled with a perfect fluid. Then the field equations of $f(T,B)$ gravity \eqref{eq:5} for the metric \eqref{eq:7} and tetrad \eqref{eq:6} can be derived as, 
\begin{eqnarray}
    -3H^2(3f_B+2f_T)+3H \dot{f_B}-3\dot{H}f_B+\frac{1}{2}f(T,B)=\kappa^2\rho \label{eq:8}\\
    -3H^2(3f_B+2f_T)-\dot{H}(3f_B+2f_T)-2H\dot{f_T}+\ddot{f_B}+\frac{1}{2}f(T,B)=-\kappa^2 p\,. \label{eq:9}
\end{eqnarray}

In Eqs.~\eqref{eq:8}--\eqref{eq:9}, $H$ be the Hubble parameter, $\rho$ and $p$ are respectively denote the  energy density and pressure of the matter component. We can calculate the Ricci scalar $R=-T+B=6(2H^2+\dot{H})$, where the torsion scalar and the boundary term are $T=6H^2$ and $B=6(\dot{H}+3H^2)$ respectively. To better understand the contributions of the modified Lagrangian, we consider $f(T,B)$ gravity as an effective fluid that appears alongside TEGR through a Lagrangian mapping $f(T,B) \rightarrow -T+\Tilde{f}(T,B)$. Using this arrangement, the Friedmann equations can be expressed as,\\
\begin{eqnarray}
    3H^{2}&=&\kappa^{2}\left(\rho+\rho_{eff}\right)\,,\\ \label{eq:10}
    3H^{2}+2\dot{H}&=&-\kappa^{2}\left(p+p_{eff}\right)\,.\label{eq:11}
\end{eqnarray} 
Where the effective energy density and effective pressure takes the form as follow,
\begin{eqnarray}
    3H^2(3\Tilde{f}_B+2\Tilde{f}_T)-3H \dot{\Tilde{f}}_B+3\dot{H}\Tilde{f}_B-\frac{1}{2}\Tilde{f}(T,B)=\kappa^2\rho_{eff} \,, \label{eq:12}\\
    -3H^2(3\Tilde{f}_B+2\Tilde{f}_T)-\dot{H}(3\Tilde{f}_B+2\Tilde{f}_T)-2H\dot{\Tilde{f}}_T+\ddot{\Tilde{f}}_B+\frac{1}{2}\Tilde{f}(T,B)=\kappa^2 p_{eff}\,.\label{eq:13}
\end{eqnarray}
The general expression for equation of state (EoS) parameter for effective fluid can be written as,
\begin{equation}
    \omega_{eff}=-1+\frac{\ddot{\Tilde{f}}_{B}-3H\dot{\Tilde{f}}_{B}-2\dot{H}\Tilde{f}_{T}-2H\dot{\Tilde{f}}_{T}}{3H^{2}(3\Tilde{f}_{B}+2\Tilde{f}_{T})-3H\dot{\Tilde{f}}_{B}+3\dot{H}\Tilde{f}_{B}-\frac{1}{2}\Tilde{f}(T,B)}\,.\label{eq:14}
\end{equation}
Further, we consider the $\tilde{f}(T,B)=\alpha T+\beta B^n$ model, such that $\tilde{f}_T=\alpha$ and $\tilde{f}_B=n\beta B^{n-1}$, where $\alpha$, $\beta$ and $n$ are positive real constants. The motivation for choosing such a form of $\tilde{f}(T,B)$, that contains the higher power of the boundary term $B$, is to understand the late time evolution of the Universe, when $n\neq 1$. In this paper, we wish to specifically focus on the change in the evolutionary behaviour of the Universe pertaining to the dynamical parameters and the stability of the models to be constructed with a small change in the value of $n$. In addition, we may have the flexibility to adjust theoretically the value of equation of state and other parameters that depends on the exponent $n$, with the result of several cosmological observations.

It can be observed from Eqs.~\eqref{eq:12}-\eqref{eq:13} that the number of unknowns are more than two viz. $H$, $p_{eff}$, $\rho_{eff}$ and obtaining an exact solution become cumbersome, therefore we need an assumed condition to obtain the solution to the field equations. We have assumed a scale factor to obtain the effective pressure, effective energy density and other parameters. However, a relationship between the matter terms can also be considered to derived the solution. Now, the Hubble parameter can be expressed as, $H=\frac{\dot{a}}{a}$, where $a$ is the scale factor. The subsequent sections contain the further derivations in two different models, one with the power law cosmology and the other with exponential scale factor.

\section{Model with Power Law cosmology}\label{sec:power_law_cosmo}

Based on the big bang theory and the inflationary scenario, present universe can be well explained by the standard cosmological model \cite{Clifton12,Linde82,Guth81}. Similar to the cosmological constant problem, the problem of the source of the cosmological constant within the standard cosmological model continues to remain unknown. The reason behind this is that the energy density of the vacuum is 120 orders of magnitude smaller than its value at the Planck time is still to explain. So within the scope of present cosmological data available and the limitations in standard model, an alternative approach has been inevitable. Power law models for late time cosmology are one example of these alternative approaches which may address the cosmological constant problem. The motivation behind this alternative approach is that (i) it has no age problem, since it can accommodate high red-shift objects; (ii) it does not encounter with the flatness and horizon problem and ; (iii) its ability to fit cosmological data. Therefore, here we shall study the behaviour of the universe with the power law cosmology corresponding to different phase of the cosmic evolution. The power law scale factor can be considered as \cite{Bahamonde18},
\begin{equation}\label{eq:15}
    a(t)=\left(\frac{t}{t_0}\right)^h\,,
\end{equation}
where $t_{0}$ is a fiducial time and $h>0$ be an arbitrary dimensionless parameter. The significance of $h$ is that it provides the evolution of standard fluid \textit{e.g.}, when $h=\frac{1}{2}$, it leads to the radiation dominated era. The torsion scalar, $T=\frac{6h^2}{t^2}$ and the boundary term $B=\frac{6h(3h-1)}{t^2}$. Now, without loss of generality, we take $\kappa=1$ and derive the effective pressure and effective energy density from  Eqs.~\eqref{eq:12}--\eqref{eq:13} respectively as,
\begin{eqnarray}
    p_{eff}&=&\left(\frac{6 h (3 h-1)}{t^2}\right)^{n-1} \left(-\frac{9 \beta  h^2 n}{t^2}+\frac{3 \beta  h n}{t^2}+\frac{\beta  (6 h (3 h-1))}{2 t^2}+\frac{2 \beta  n (n-1) (2 n-1)}{t^2}\right)-\left(\frac{3 \alpha  h^2}{t^2}-\frac{2 \alpha  h}{t^2}\right)\,, \label{eq:16}\\
    \rho_{eff}&=&\left(\frac{6 h (3 h-1)}{t^2}\right)^{n-1} \left(\frac{9 \beta  h^2 n}{t^2}-\frac{3 \beta  h n}{t^2}+\frac{6 \beta  h n (n-1)}{t^2}-\frac{\beta  (6 h (3 h-1))}{2 t^2}\right)+\left(\frac{3 \alpha  h^2}{t^2}\right)\,. \label{eq:17}
\end{eqnarray}

\begin{figure}[!hbt]
\centering
\includegraphics[width=80mm]{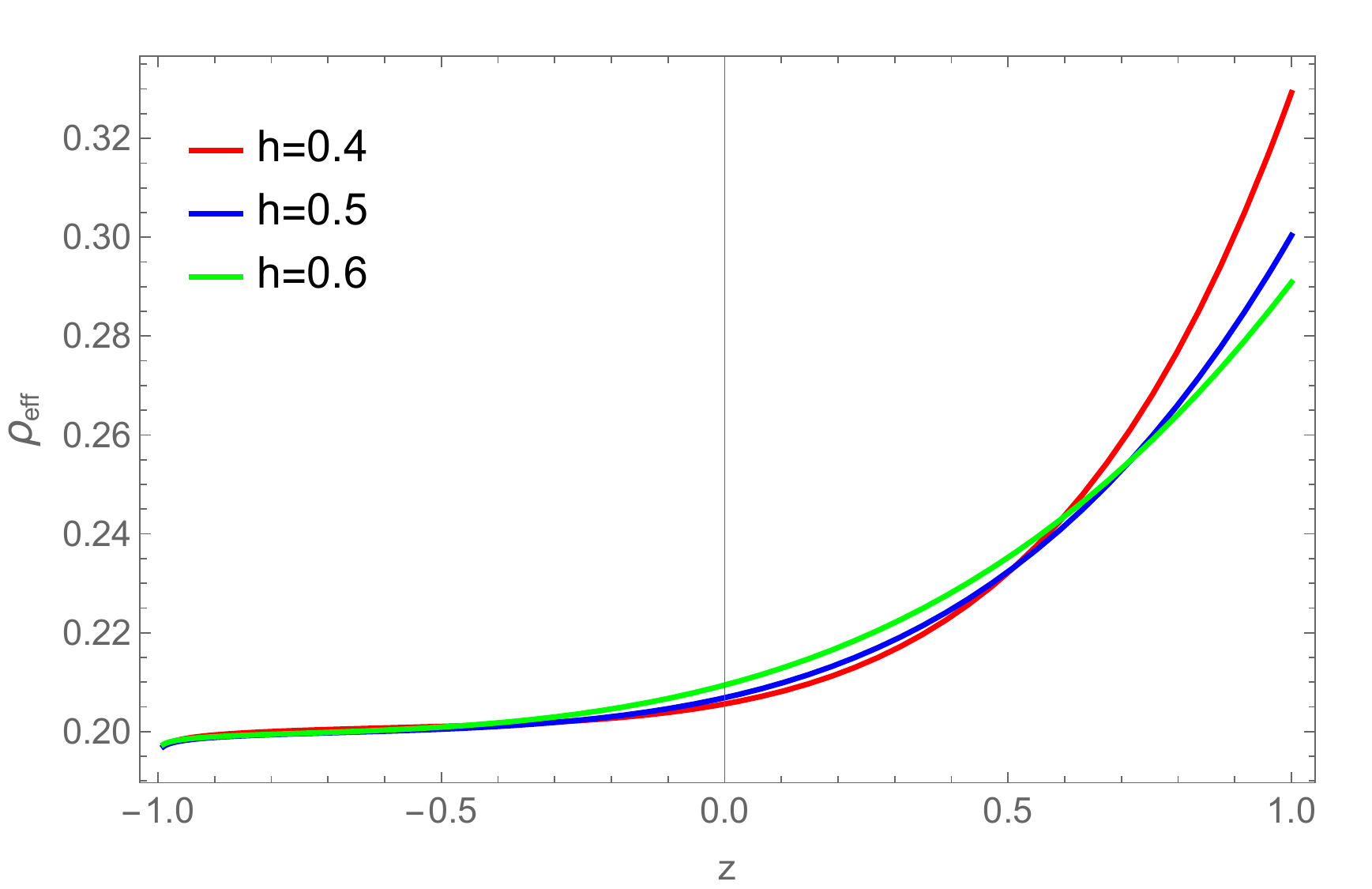}
\includegraphics[width=80mm]{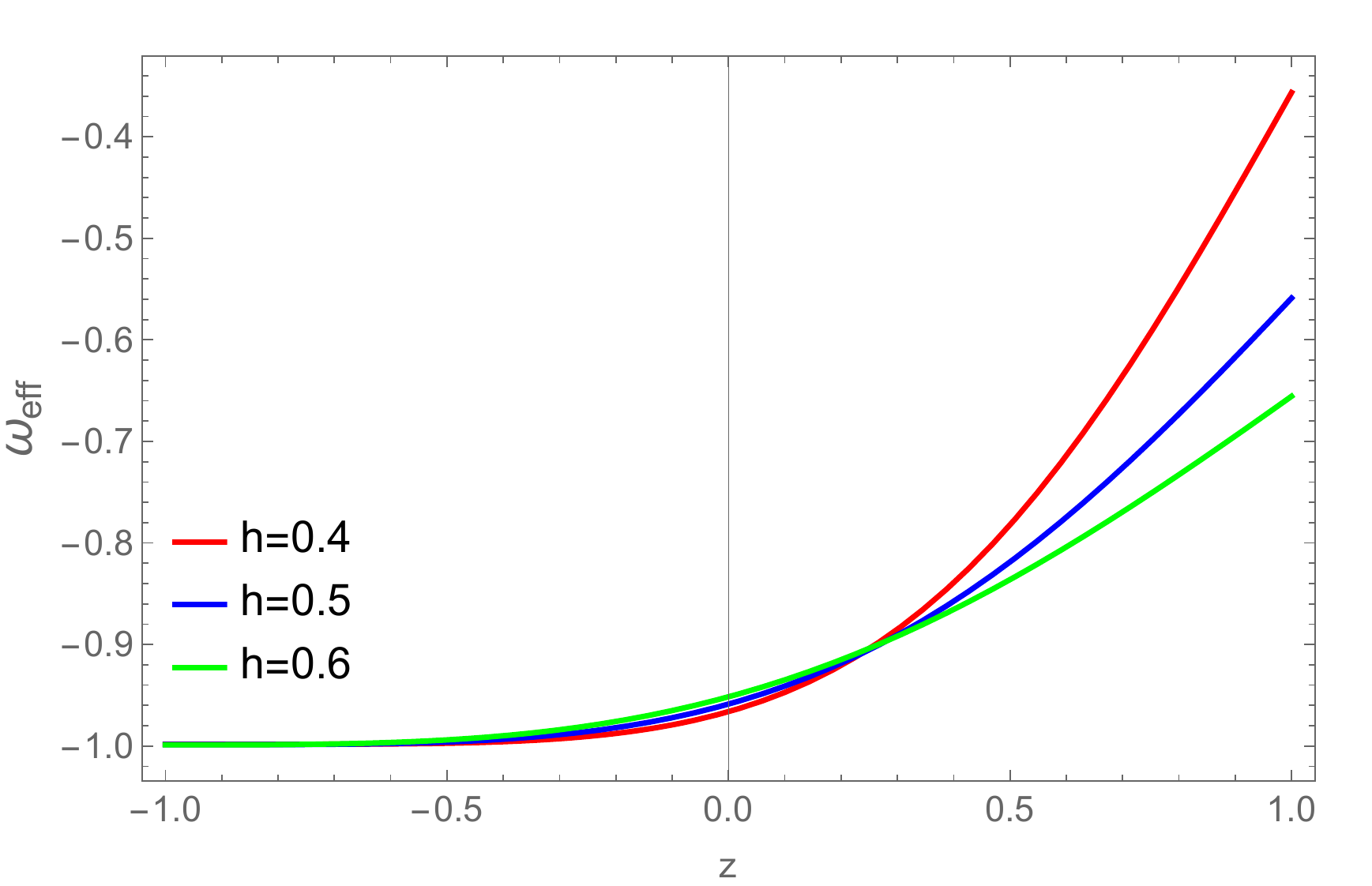}
\caption{Effective energy density (left panel) and effective EoS parameter (right panel) vs redshift for varying $h$. The other parameter's values are, $\alpha=0.01$, $\beta=-0.4$, $n=0.001$, $t_0=1.1$.} 
\label{Fig1}
\end{figure}

\begin{figure}[!hbt]
\centering
\includegraphics[width=80mm]{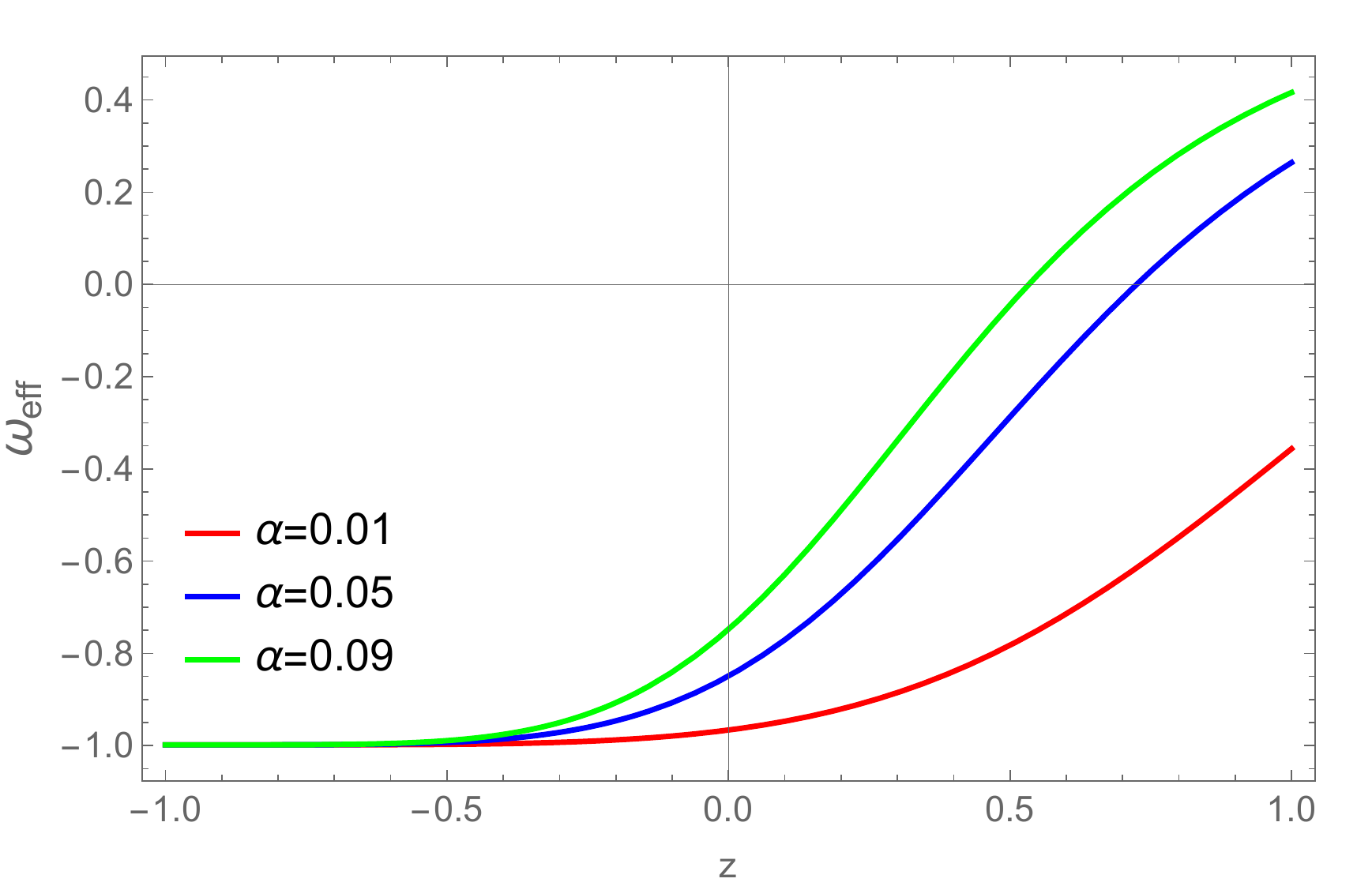}
\includegraphics[width=80mm]{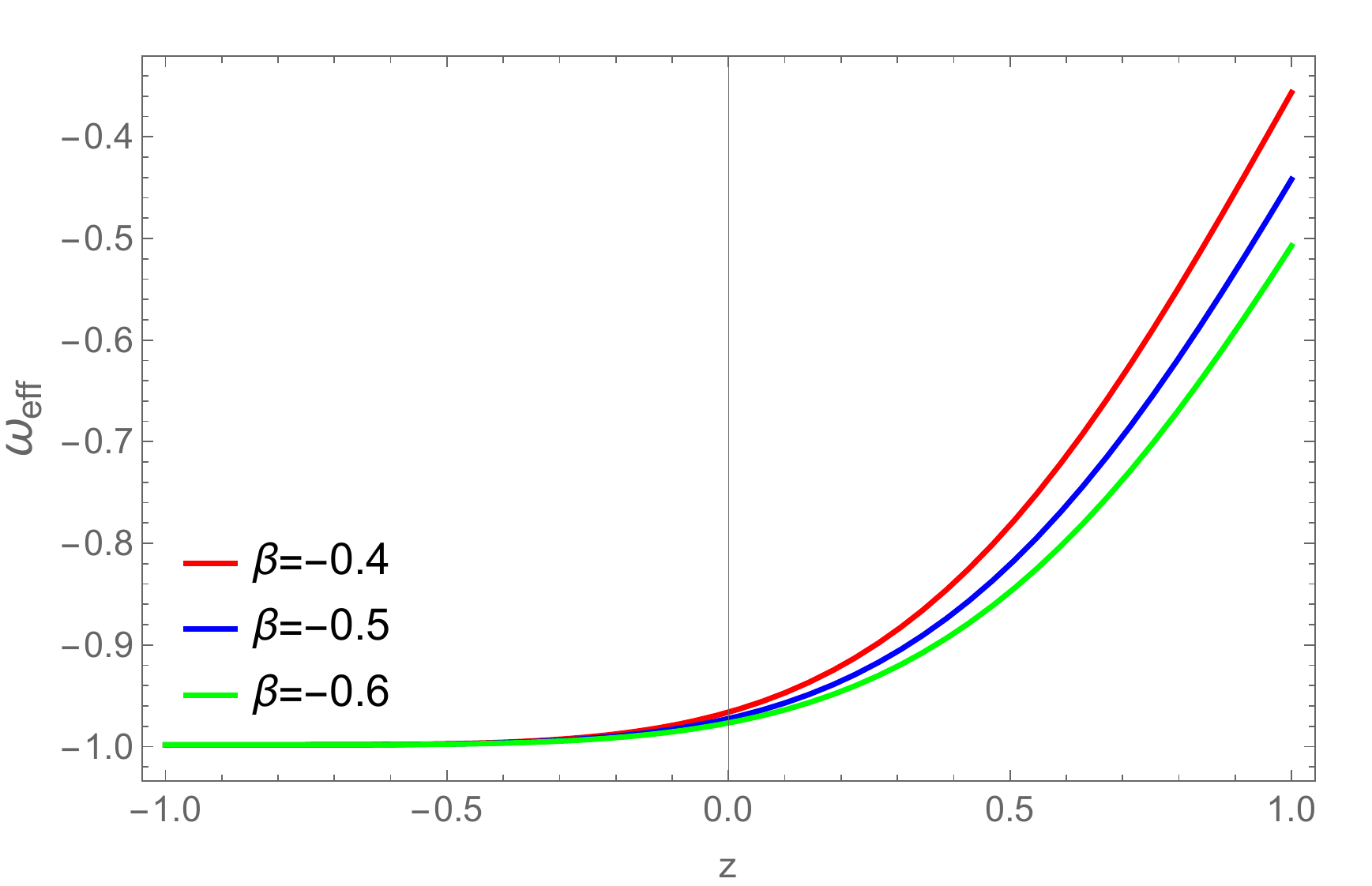}
\caption{The EoS parameter with varying $\alpha$ vs redshift with $h=0.4$, $\beta=-0.4$, $n=0.001$, $t_0=1.1$  (left panel) and varying $\beta$(right panel) vs redshift for varying $\beta$ with $h=0.4$, $\alpha=0.001$, $n=0.001$, $t_0=1.1$.}
\label{Fig2}
\end{figure}

The equation of state parameter (EoS) of the perfect fluid has been characterized by a dimensionless number $\omega=\frac{p}{\rho}$. We are intending to study the behaviour of the expanding universe through this cosmological model and in an expanding universe fluids with larger EoS disappear more quickly as compared to the fluid with smaller EoS. An important outcome of the cosmological observations is the measurement of EoS parameter i.e. they would have around early big bang with curvature, between $\omega=-\frac{1}{3}$ and $\omega=0$. By measuring $\omega$, we can distinguish the cosmological constant from the quintessence. In addition, the accelerated expansion of the universe can be characterized by the EoS parameter, $\omega=-1$, leads to the cosmological constant. Therefore, we derive the effective EoS parameter from  \eqref{eq:16}--\eqref{eq:17} as,
\begin{equation}
\omega_{eff}=-\frac{3 \alpha  h^2-2 \alpha  h+\beta  6^{n-1} (n-1) (3 h+2 n-1) (3 h-2 n) \left(\frac{h (3 h-1)}{t^2}\right)^{n-1}}{3 \alpha  h^2+\beta  h 2^{n-1} 3^n (n-1) (3 h+2 n-1) \left(\frac{h (3 h-1)}{t^2}\right)^{n-1}}\,. \label{eq:18}
\end{equation}
We have shown the evolutionary behaviour of effective energy density (left panel) and effective EoS parameter (right panel) in FIG.~\ref{Fig1}. First we checked the behaviour with varying value of $h$. The energy density remains entirely in the positive region. The EoS parameter evolve from a lower negative value i.e. from the quintessence phase and approaches asymptotically to $-1$ at late time of the evolution. So, it shows the $\Lambda$CDM behaviour at late time of the evolution. It has been observed that lower $h$ value decreases more rapidly as compared to the higher $h$ value. At the same time, we are interested to access the behaviour of the EoS parameter with the changed value of the parameter $\alpha$ and $\beta$ described in the functional form of $\tilde{f}(T,B)$. So, in FIG.~\ref{Fig2}, we have seen that with varying $\alpha$ and $\beta$, the EoS parameter have similar behaviour of approaching $\Lambda$CDM phase at late time. However, with varying $\beta$, the model approaches to $-1$ more rapidly as compared to the varying $\alpha$. The reason could be the choice of the functional of $\tilde{f}(T,B)$, which is independent of torsion $T$ in the first derivative. When $n=1$, i.e. in the linear form of $\tilde{f}(T,B)$, Eq. \eqref{eq:18} reduces to $\omega=-1+\frac{2}{3h}$ and the $\Lambda$CDM model can be realised for the higher value of $h$. Also, when $h=\frac{2}{3}$, it leads to the dust model. The advantage of considering a value other than $n=1$ resulted in obtaining $\Lambda$CDM behaviour immediately after the present time with the present value of $\omega_0$ lie in the range of the value provided by cosmological observations. Also, the evolutionary behaviour of the EoS parameter has been studied with varying values of the model and scale factor parameter to assess its effect. Though at late time the evolutionary behaviour remains alike for all the parameters, but at early time the evolution begins from different phases for different parameters.    


We shall now give some analysis of the model with the energy conditions, scalar field reconstruction and the stability behaviour. 
\subsection{Energy Conditions}\label{sec:Energy Condition}

Energy conditions are mathematically framed boundary conditions to keep the energy density positive, however it is not the physical constraint on the model. The energy conditions stipulate as: (i) Null Energy Condition (NEC), $\rho_{eff}+p_{eff} \geq 0$; (ii) Weak Energy Condition(WEC), $\rho_{eff}+p_{eff} \geq 0,$ $\rho_{eff} \geq 0$; (iii) Strong Energy Condition(SEC): $\rho_{eff}+3p_{eff}\geq0$, and (iv) Dominant Energy Condition(DEC): $\rho_{eff}-p_{eff}\geq0$. In the context of dark energy, an anti gravity leads to the negative pressure and hence it is expected that the SEC should violate and to note even in the context of perfect fluid SEC does not imply to WEC. The SEC is used in the classic Hawking-Penrose singularity theorem, whose violation allows for the observed accelerated expansion \cite{Hawking73}. The NEC is sufficient to ensure that the universe density decreases as its size grows. The SEC suggests that the universe is slowing down, and this result remains true regardless of whether the universe is open, flat, or closed \cite{Matt97}. We can express the energy conditions of the model using Eqs.~\eqref{eq:16} and \eqref{eq:17} as,
\begin{eqnarray}
\rho_{eff}+p_{eff}&=&\frac{\beta  2^n 3^{n-1} (n-1) n (3 h+2 n-1) \left(\frac{h (3 h-1)}{t^2}\right)^{n-1}}{t^2}+\frac{2 \alpha  h}{t^2}\,, \nonumber \\
\rho_{eff}+3p_{eff}&=&-\frac{\beta  6^n (n-1) (h-n) (3 h+2 n-1) \left(\frac{h (3 h-1)}{t^2}\right)^{n-1}}{t^2}-\frac{6 \alpha  (h-1) h}{t^2}\,, \nonumber \\
\rho_{eff}-p_{eff}&=&\frac{2 \alpha  h (3 h-1)}{t^2}-\frac{\beta  2^n 3^{n-1} (n-1) (n-3 h) (3 h+2 n-1) \left(\frac{h (3 h-1)}{t^2}\right)^{n-1}}{t^2}.\label{eq:13a} 
\end{eqnarray}

\begin{figure}[!hbt]
\centering
\includegraphics[width=55mm]{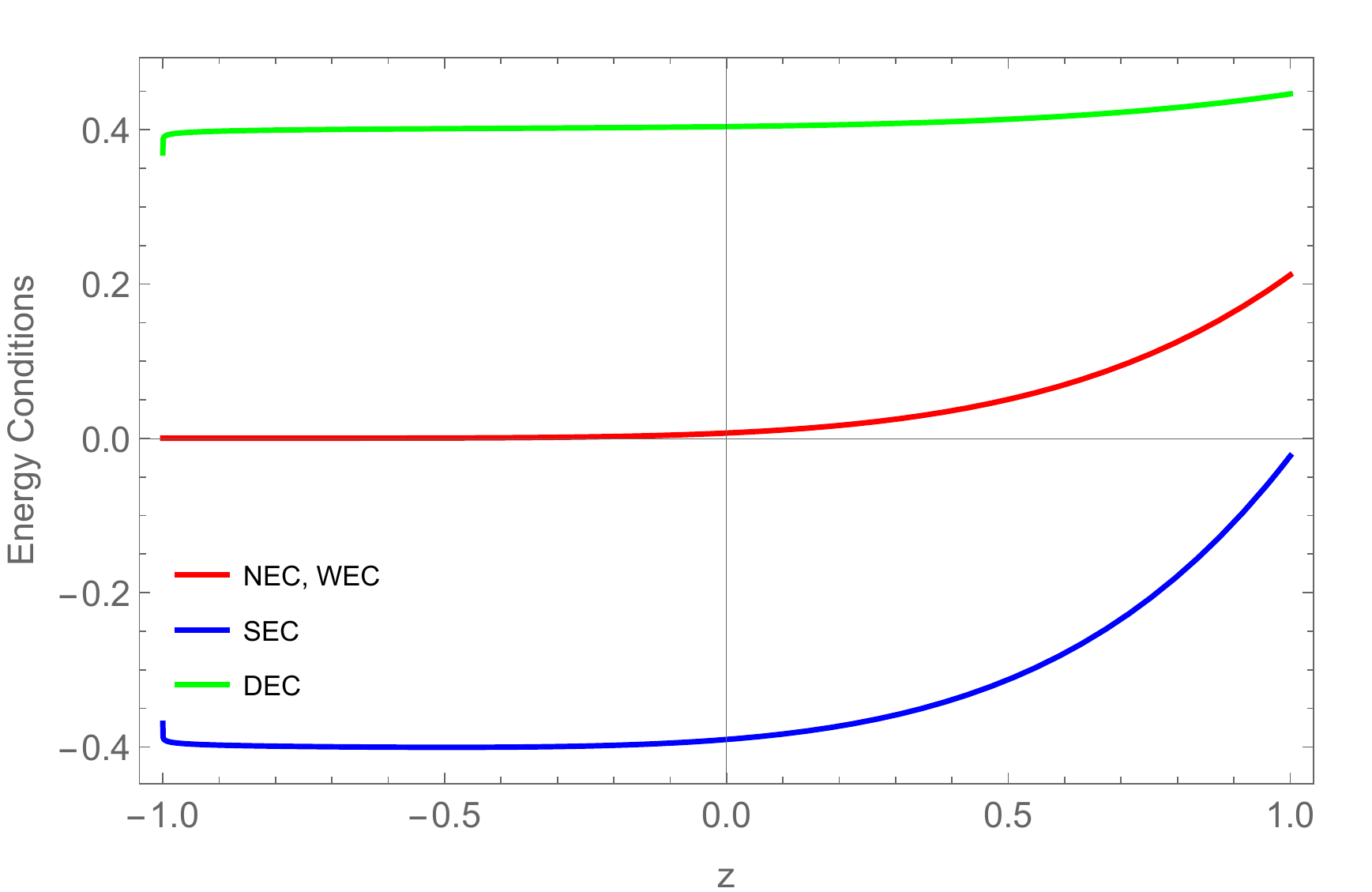}
\includegraphics[width=55mm]{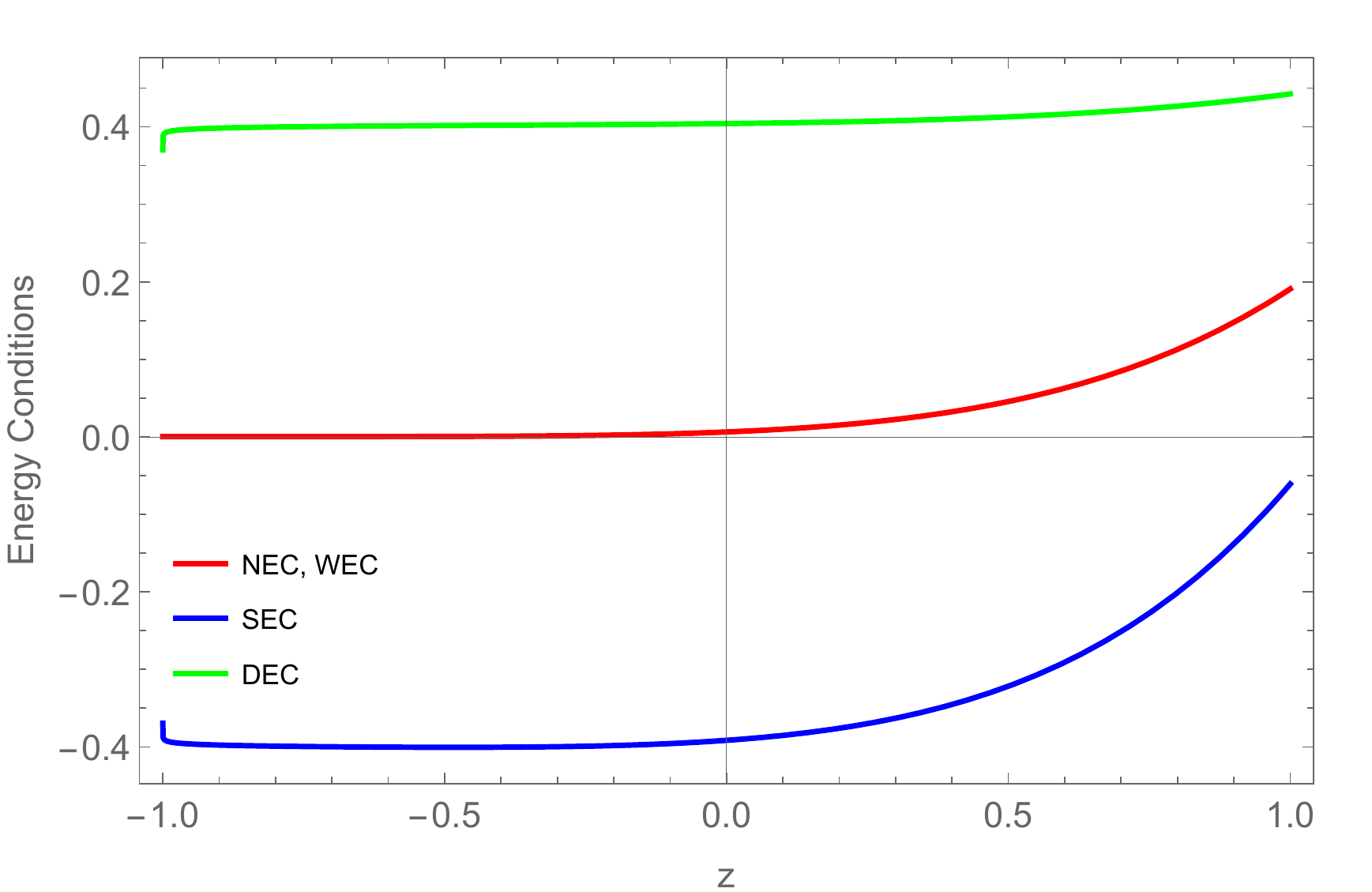}
\includegraphics[width=55mm]{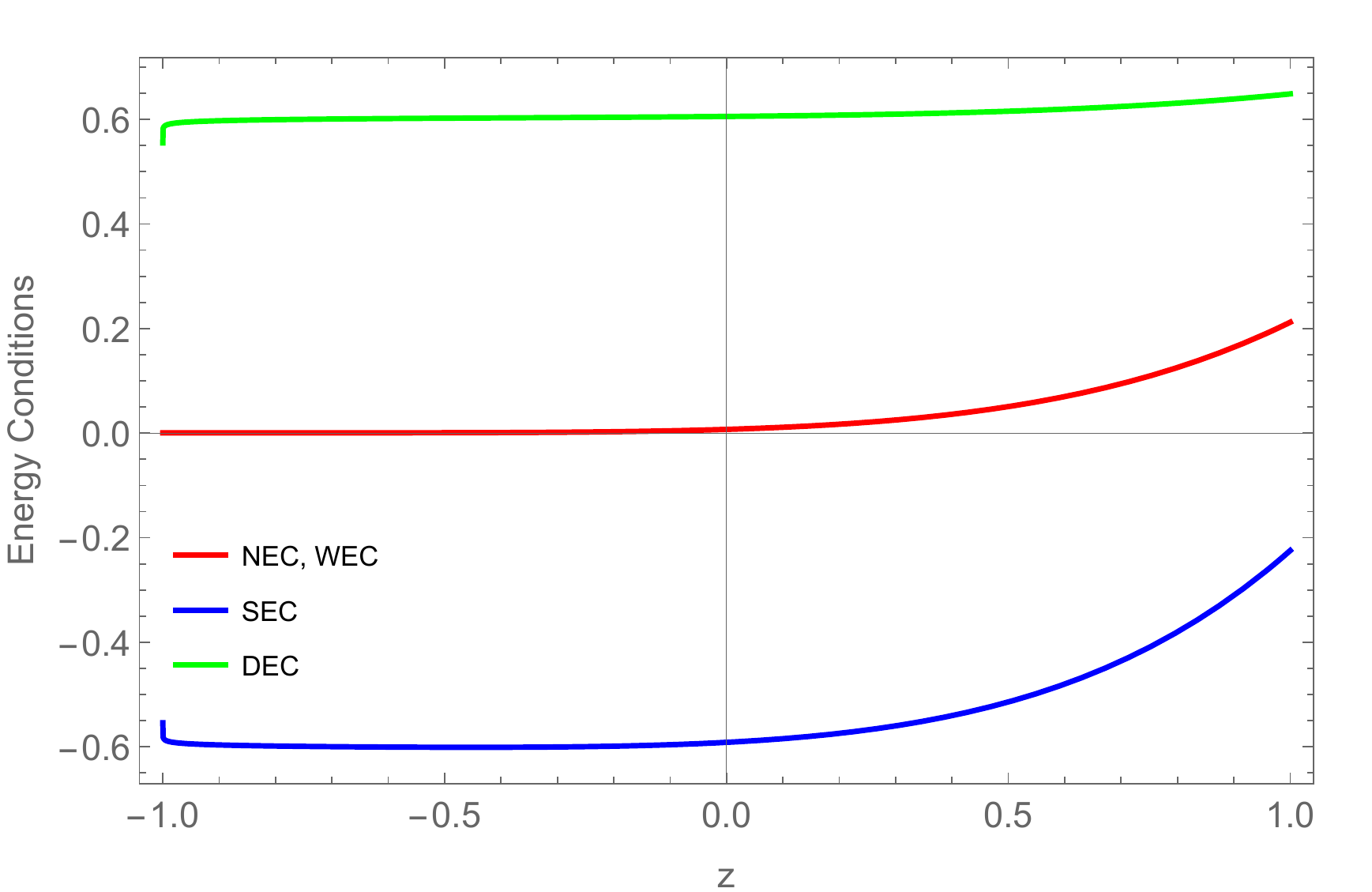}
\caption{Energy conditions vs redshift with varying $\alpha$ and $\beta$ as $\alpha=0.01, \beta=-0.4$ (left panel), $\alpha=0.009, \beta=-0.4$ (middle panel), $\alpha=0.01, \beta=-0.6$(right panel). The other parameters value are, $h=0.4$, $n=0.001$, $t_0=1.1$.}
\label{Fig3}
\end{figure}

In Fig.~\ref{Fig3}, we have presented three plots to assess the behaviour of the energy conditions with varying $\alpha$ and $\beta$ value, which are the two parameters of the functional $\tilde{f}(T,B)$. The motivation behind multiple graphs is to check if there is any change in the behaviour of violation of SEC as required in modified theories of gravity. It can be observed in all three plots that the SEC remain in the negative region, hence confirms the violation as required in the modified theories of gravity. The NEC at the initial stage does not violate and at late phase vanishes. Another observation is that with varying $\alpha$ (left panel, middle panel), the behaviour of the energy conditions remain same except the fact that the SEC changes more rapidly in higher value of $\alpha$. At the same time with varying $\beta$ (left panel, right panel) the behaviours are remain same with lower value of $\beta$ increases slowly than its higher value.

\subsection{Scalar Field Reconstruction}\label{sec:Scalar Field Reconstruction}

Scalar field has been instrumental in modified theories of gravity with slow varying potential to describe the dark energy and inflation scenario. In several cosmological sense, like inflation, the scalar field has been used and can be constrained through the CMB observations. Scalar fields arise in low-energy limit of higher dimensional theories and can mimic the evolution of matter. Moreover, it does not require the fine tuning either in its initial conditions or parameter in order to change the accelerated expansion behaviour at late time. This is because of the change in the cosmological equation of state when a non relativistic dark matter component, directly coupled to the scalar field, begins to make a significant contribution to the total density. We reconstructed the scalar filed and analyse the behaviour of scalar field and self interacting potential in the contest of modified theories of gravity with respect to redshift $z$. Now, the EoS parameter can also be expressed as, $\omega=\frac{p_{\phi}}{\rho_{\phi}}=\frac{\frac{\dot{\phi}^2}{2}-V(\phi)}{\frac{\dot{\phi}^2}{2}+V(\phi)}$, where $\phi$ and $V(\phi)$ are respectively the scalar field and self interacting potential. In the previous section, we have seen that at late time the model approaches to $\Lambda$CDM, hence using $\omega=-1$, we can observed that the scalar field remains constant at late time. Hence, the late time cosmic acceleration phenomena can also be modelled through the use of scalar field. We can obtain the pressure and energy density with Friedmann background as,
\begin{eqnarray}
    p_{\phi}=\epsilon \frac{\dot{\phi}^2}{2}-V(\phi) \,,\label{eq:14}\\
    \rho_{\phi}=\epsilon \frac{\dot{\phi}^2}{2}+V(\phi)\,.
    \label{eq:15}
\end{eqnarray} 
where $\epsilon=-1$ and $\epsilon=+1$  respectively represents for phantom and quintessence field. We can reconstruct the model with the scalar field with the following expressions.
\begin{eqnarray}
    \epsilon \dot{\phi}^2&=&\frac{\beta  2^n 3^{n-1} (n-1) n (3 h+2 n-1) \left(\frac{h (3 h-1)}{t^2}\right)^{n-1}}{t^2}+\frac{2 \alpha  h}{t^2}\,,\\
    2V(\phi)&=& \frac{2 \alpha  h (3 h-1)}{t^2}-\frac{\beta  2^n 3^{n-1} (n-1) (n-3 h) (3 h+2 n-1) \left(\frac{h (3 h-1)}{t^2}\right)^{n-1}}{t^2}\,.
\end{eqnarray}

\begin{figure}[!hbt]
\centering
\includegraphics[width=80mm]{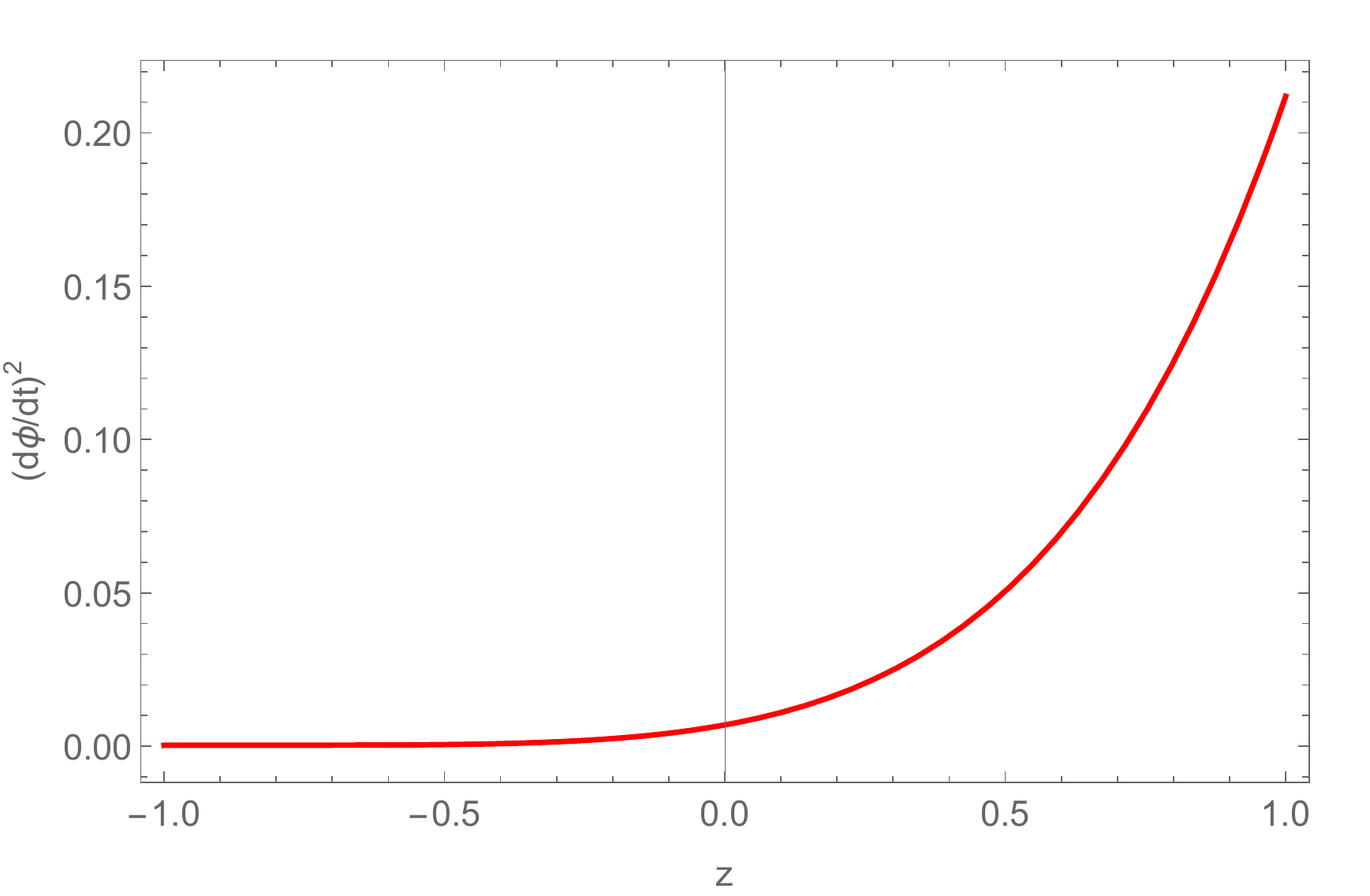}
\includegraphics[width=80mm]{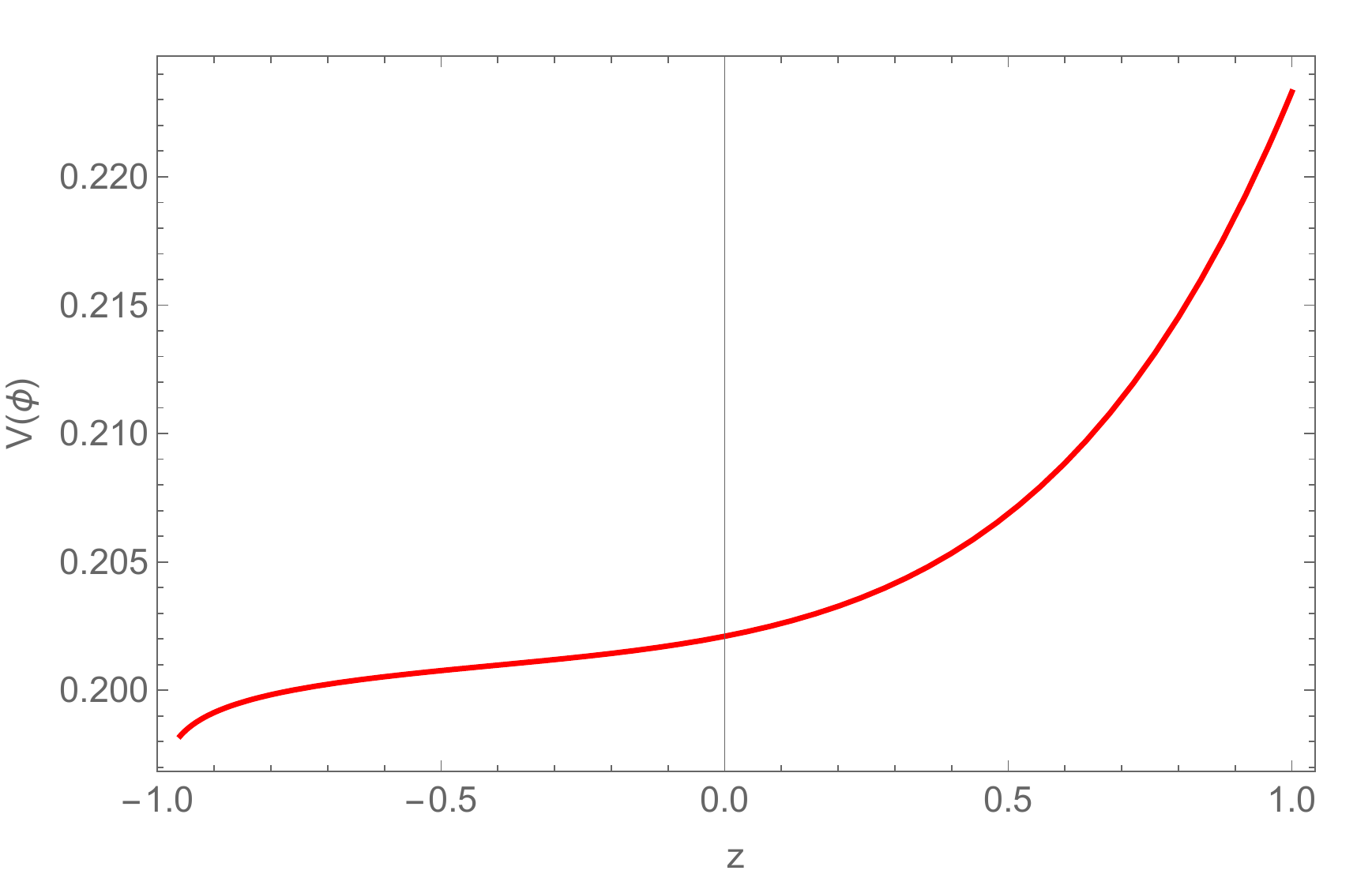}
\caption{Squared slope of reconstructed scalar field as a function of redshift model $\dot{\phi}^2$ (left panel) and $V(\phi)$ (right panel) vs redshift with $h=0.4$, $\alpha=0.01$,  $\beta=-0.4$, $n=0.001$, $t_0=1.1$.} 
\label{Fig4}
\end{figure}

In Fig.~\ref{Fig4}, the squared slope of reconstructed scalar field shows decreasing behaviour from early to late time and ultimately vanishes. The scalar potential also decreases from higher value and gradually decreases over the time, however maintains in the positive profile.  The behaviour of the scalar field is model dependent and more information can be obtained through the behaviour of the Hubble rate.

\subsection{Stability Analysis of the Model}\label{sec:Stability analysis}

To analyse the dynamical behaviour of the cosmological models  in modified theories of gravity, the governing equations remain highly non-linear. Several assumptions are made to solve the system. But the degree of generality of these assumptions is difficult to assess, hence there is a need to test the qualitative properties of the field equations, which can be performed through stability analysis \cite{Charters01}. Mechanical stability of the cosmic fluid reveals the stability of the model which can be obtained by calculating the adiabatic speed of sound through the cosmic fluid as, $C_s^2=\frac{dp}{d\rho}=\frac{dp/dt}{d\rho/dt}$, where $C_s^2$ is measured in the unit of the square of the speed of light in vacuum \cite{Balbi07,Xu13}. The model is stable if $C_s^2>0$ and unstable for $C_s^2<0$. We can calculate the stability function using Eqs.~\eqref{eq:16} and \eqref{eq:17} and can be derived as,
\begin{equation}
\frac{dp_{eff}}{d\rho_{eff}}=\frac{\beta  6^n (n-1) n t^2 (2 n-3 h) (3 h+2 n-1) \left(\frac{h (3 h-1)}{t^2}\right)^n-6 \alpha  h^2 (9 (h-1) h+2)}{3 t^3 \left(\frac{6 \alpha  h^3 (3 h-1)}{t^3}+\frac{\beta  h 6^n (n-1) n (3 h+2 n-1) \left(\frac{h (3 h-1)}{t^2}\right)^n}{t}\right)}\,.
\end{equation}

\begin{figure}[!htp]
\centering
\includegraphics[width=80mm]{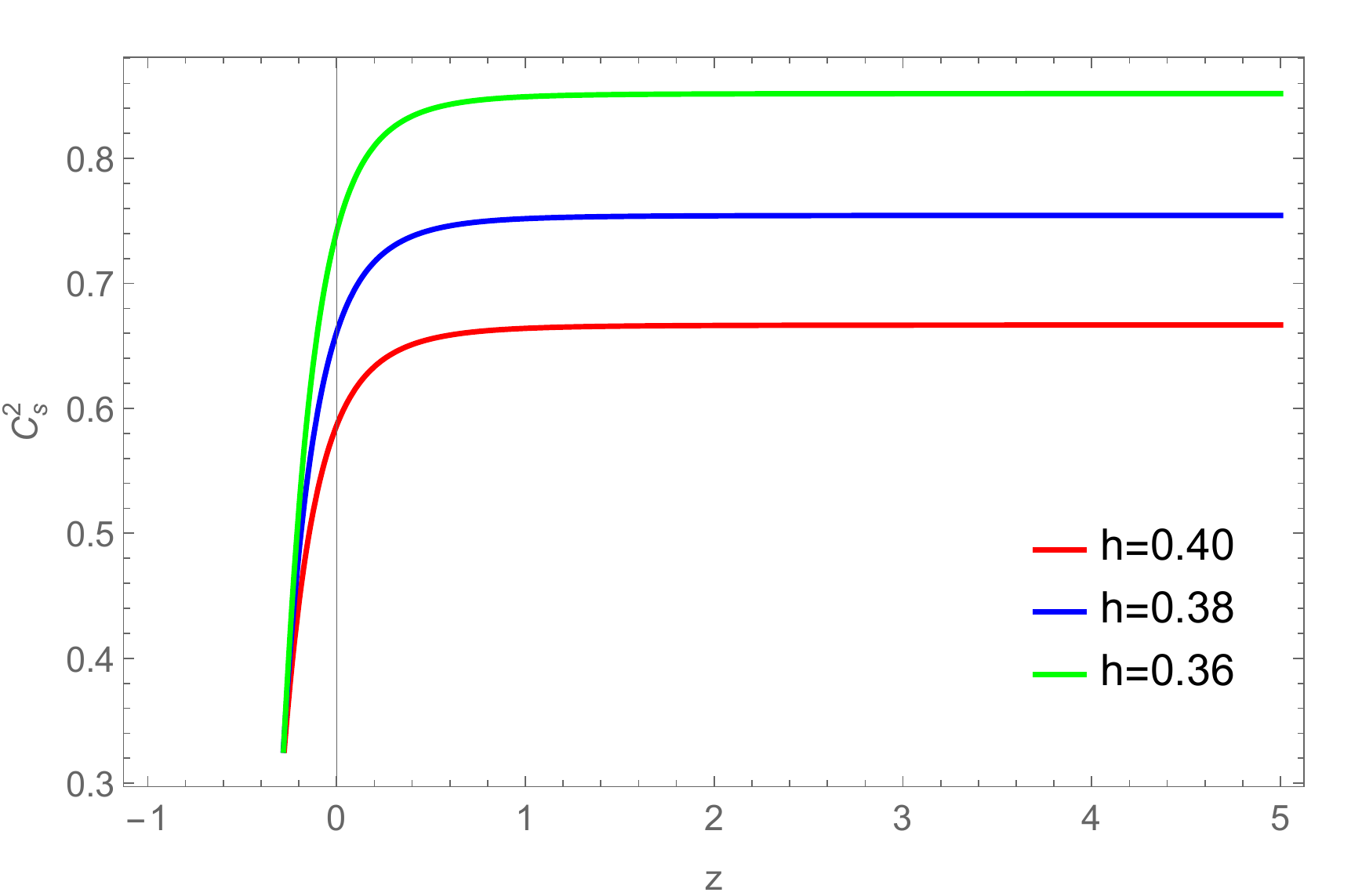}
\caption{Stability as a function of redshift for the parametric value, $\alpha=0.01$, $\beta=-0.4$, $n=0.001$, $t_0=1.1$.}
\label{Fig5}
\end{figure}

Sharif and Ikram \cite{Sharif17} have performed the stability analysis of reconstructed $f(G,T)$ cosmological model. Using dynamical system analysis, Shah and Samanta \cite{Shah19} presented the stability of $f(R)$ cosmological model. Franco et al. \cite{Franco20} have checked the stability of the cosmological model in $f(T,B)$ gravity.  Mishra et al. have performed the stability analysis of the cosmological model in two fluid scenario \cite{Mishra21}. The graphical behaviour has been given in Fig.~\ref{Fig5} for already chosen  appropriate value of the model and scale factor parameters with varying $h$. In all the values, the model obtained to be stable. It is to note here that, with the other combinations of the values of the parameters, the stability behaviour remains unchanged. Therefore, we can say that the methodology adopted to solve the system and to frame the cosmological model is obtained to be stable.  

\section{Model with exponential scale factor}\label{sec:exponential_Scale_factor}

\begin{figure}[!htp]
\centering
\includegraphics[width=80mm]{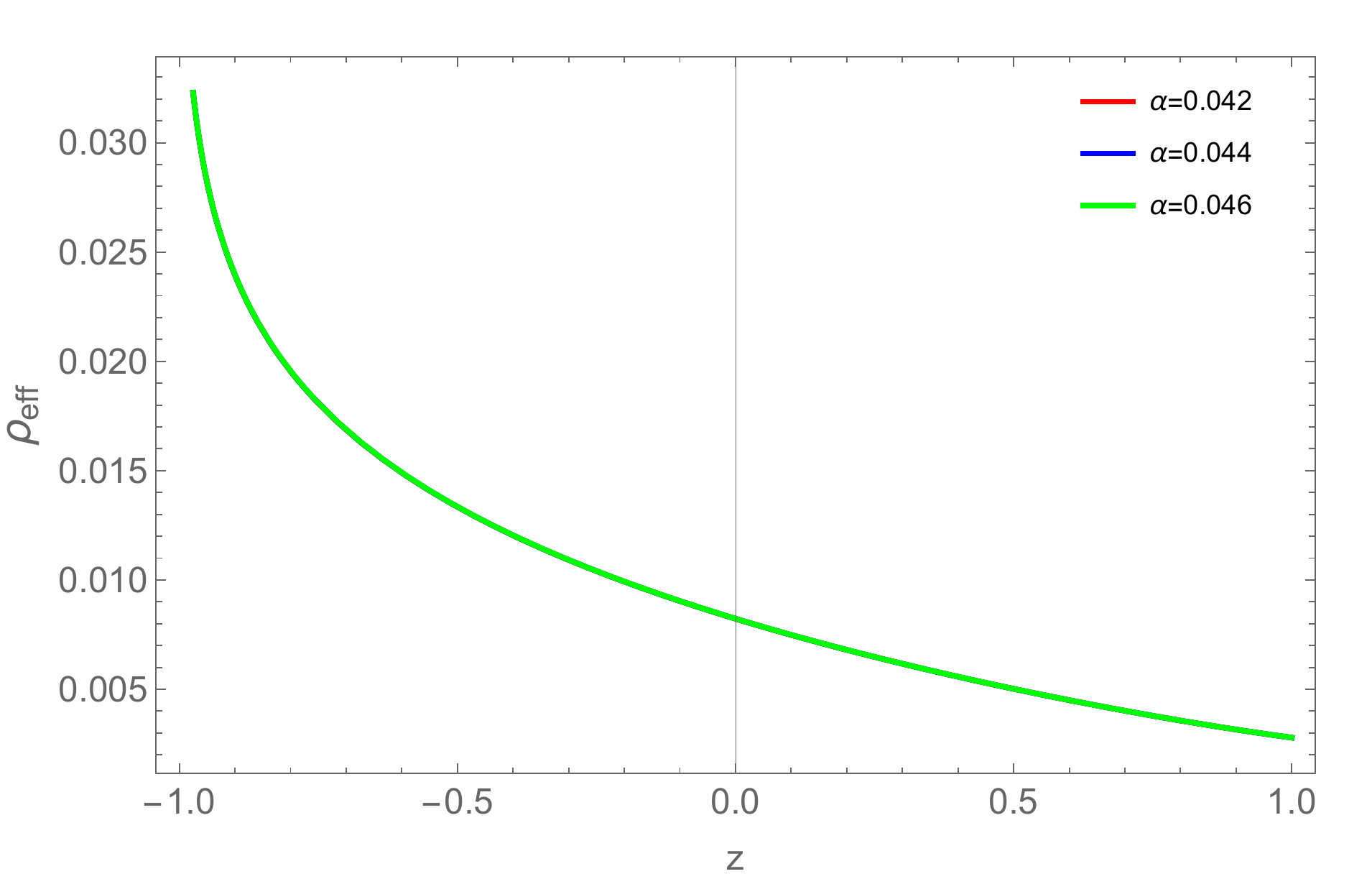}
\includegraphics[width=80mm]{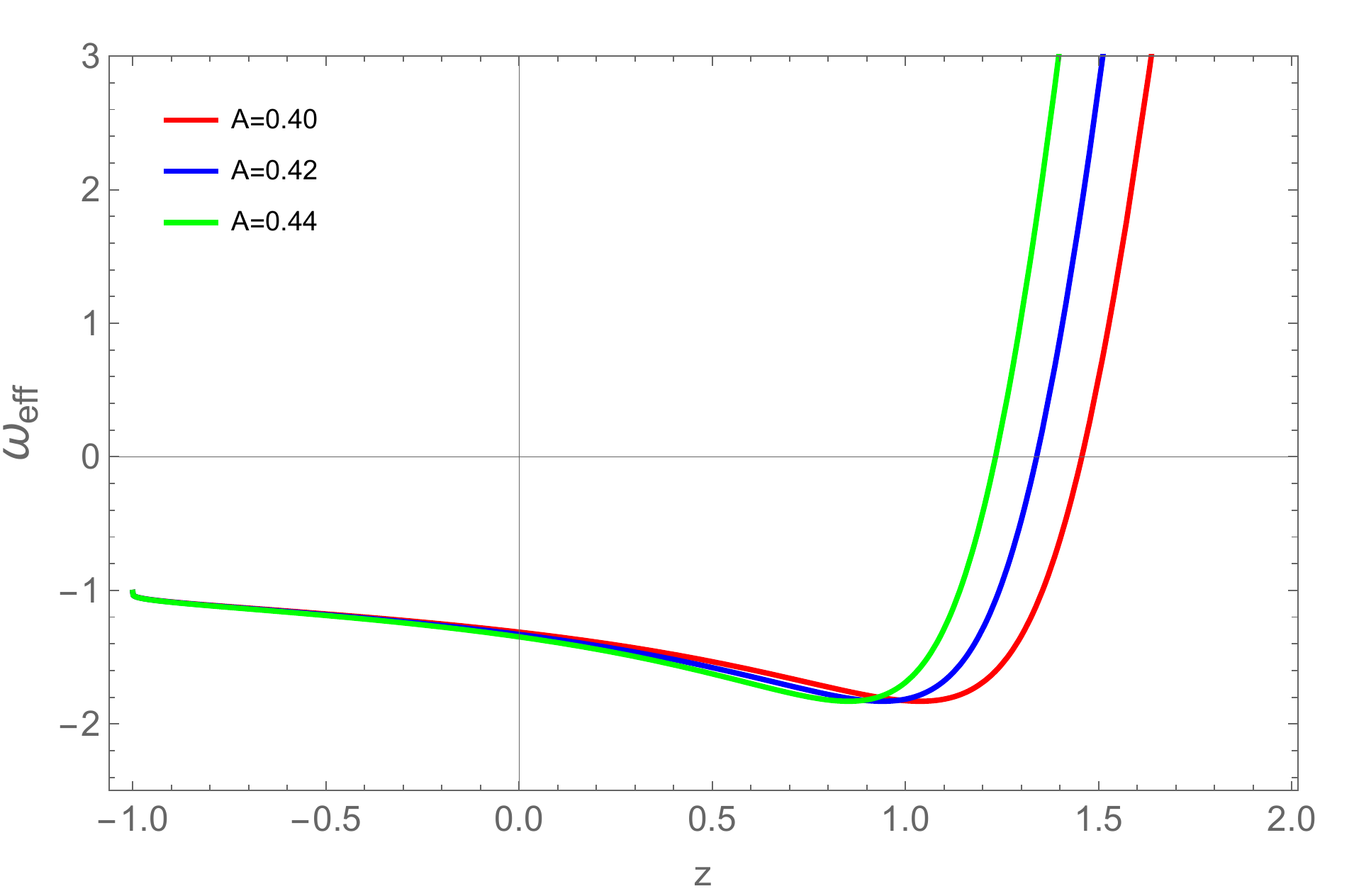}
\caption{Effective energy density (left panel) for varying values of $\alpha$ and effective EoS parameter (right panel) for varying parameter $A$ vs redshift. The other parameters value are, $\beta=-5.1$, $n=0.8$, $t_{\star}=10.1$, $\zeta=0.01$.} 
\label{Fig6}
\end{figure}

Motivated with the accelerating behaviour of the cosmological model with power law function in $f(T,B)$, here we have considered an exponential scale factor in the form, 
\begin{equation}
a(t)=A \exp\left(\zeta\frac{t^2}{t_{\star}^2}\right),
\end{equation}
where $\zeta>0,, A>0$ are constants and $t_{\star}$ represents the arbitrary time. The Hubble parameter can be calculated as $H= \frac{2 \zeta t}{t_{\star}^2}$. The  torsion scalar,  $T=\frac{24 \zeta ^2 t^2}{t_{\star}^4}$ and boundary term $B= \frac{12 \zeta  \left(t_{\star}^2+6 \zeta  t^2\right)}{t_{\star}^4}$. For brevity, we set $a(t_{0}) = 1$, and for some arbitrary time $t_{\star} >0 $. Then, we obtain the relation for $t_{0}$ as,
\begin{equation}
    t_{0}=\sqrt{\frac{-t_{\star}^2}{\zeta} ln A}\,.
\end{equation}
In this case, $A$ $\in$ (0, 1). The exponential scale factor is used in the study of bouncing cosmology, the symmetric bounce is characterized by the exponential scale factor\cite{Caruana20}. The dynamical parameters for the case, $\tilde{f}(T,B) = \alpha T + \beta B^n $ can be calculated as,
\begin{eqnarray}
p_{eff}&=&\frac{\beta  t_{\star}^4 12^n (n-1) n \chi ^n \left(t_{\star}^2+6 \zeta  (2 n-3) t^2\right)}{\left(t_{\star}^2+6 \zeta  t^2\right)^3}-\frac{12 \zeta ^2 t^2 \left(2 \alpha +3 \beta  n (12\chi)^{n-1}\right)}{t_{\star}^4}-\frac{2 \zeta  \left(2 \alpha +3 \beta  n (12\chi)^{n-1}\right)}{t_{\star}^2} \nonumber\\
& &+\frac{12 \alpha  \zeta ^2 t^2}{t_{\star}^4}+\frac{\beta  (12\chi)^n}{2} \,, \nonumber \\
\rho_{eff}&=&\frac{\beta  t_{\star}^8 12^n (n-1) \chi ^n-\beta  t_{\star}^6 \zeta  12^{n+1} (n-1)^2 t^2 \chi ^n+12 t_{\star}^4 \zeta ^2 t^2 \left(2 \alpha +3 \beta  (n-1) t^2 (12\chi)^n\right)+288 \alpha  t_{\star}^2 \zeta ^3 t^4+864 \alpha  \zeta ^4 t^6}{2 t_{\star}^4 \left(t_{\star}^2+6 \zeta  t^2\right)^2}\,,\nonumber\\
\omega_{eff}&=&\frac{-2 t_{\star}^4 \left(t_{\star}^2+6 \zeta  t^2\right)^2 \left(-\frac{\beta  t_{\star}^4 12^n (n-1) n \chi ^n \left(t_{\star}^2+6 \zeta  (2 n-3) t^2\right)}{\left(t_{\star}^2+6 \zeta  t^2\right)^3}+\frac{2 \zeta  \left(2 \alpha +3 \beta  n (12\chi)^{n-1}\right)}{t_{\star}^2}-\frac{12 \alpha  \zeta ^2 t^2}{t_{\star}^4}-\frac{\beta  (12\chi)^n}{2}\right)}{\beta  t_{\star}^8 12^n (n-1) \chi ^n-\beta  t_{\star}^6 \zeta  12^{n+1} (n-1)^2 t^2 \chi ^n+12 t_{\star}^4 \zeta ^2 t^2 \left(2 \alpha +3 \beta  (n-1) t^2 (12\chi)^n\right)+288 \alpha  t_{\star}^2 \zeta ^3 t^4+864 \alpha  \zeta ^4 t^6}\nonumber\\
& &-\frac{2  \left(t_{\star}^2+6 \zeta  t^2\right)^2 \left(12 \zeta ^2 t^2 \left(2 \alpha +3 \beta  n (12\chi)^{n-1}\right)\right)}{\beta  t_{\star}^8 12^n (n-1) \chi ^n-\beta  t_{\star}^6 \zeta  12^{n+1} (n-1)^2 t^2 \chi ^n+12 t_{\star}^4 \zeta ^2 t^2 \left(2 \alpha +3 \beta  (n-1) t^2 (12\chi)^n\right)+288 \alpha  t_{\star}^2 \zeta ^3 t^4+864 \alpha  \zeta ^4 t^6}\,.\nonumber \\
\end{eqnarray}

where $\chi =\frac{\zeta  \left(t_{\star}^2+6 \zeta  t^2\right)}{t_{\star}^4}$,
the graphs for effective energy density and effective EoS parameter with varying parametric values $\alpha$ and $A$ respectively are plotted in FIG.~\ref{Fig6}. The energy density lies in the positive region and showing similar behaviour for varying value of $\alpha$ at early time and decreases slight slowly for increasing values of $\alpha$. We have observed that all the curves for EoS parameter for varying values of $A=0.40,0.42, 0.44$ merging in a single curve and approaches $\Lambda$CDM at late time, therefore we explored its behaviour with this particular set of values of the scale factor parameter $A$. The EoS parameter shows slight shifting in the graphical behaviour for different values of $A$ at early time. The EoS parameter shows more decreasing behaviour with increasing values of $A$ and approaches to $-1$ at late time. The model coincide with $\Lambda$CDM at late time for different values of $A$. For $n=1$, we have $\omega=-1-\frac{t_{\star}^2}{3 \zeta  t^2}$. We can infer that in the case of $\tilde{f}(T,B)=\alpha T+\beta B$, the EoS parameter approaches to $\Lambda$CDM at infinitely late time, however in this form of $\tilde{f}(T,B)$ the $\Lambda$CDM behaviour obtained earlier.

\subsection{Energy Conditions}

\begin{figure}[!htp]
\centering
\includegraphics[width=80mm]{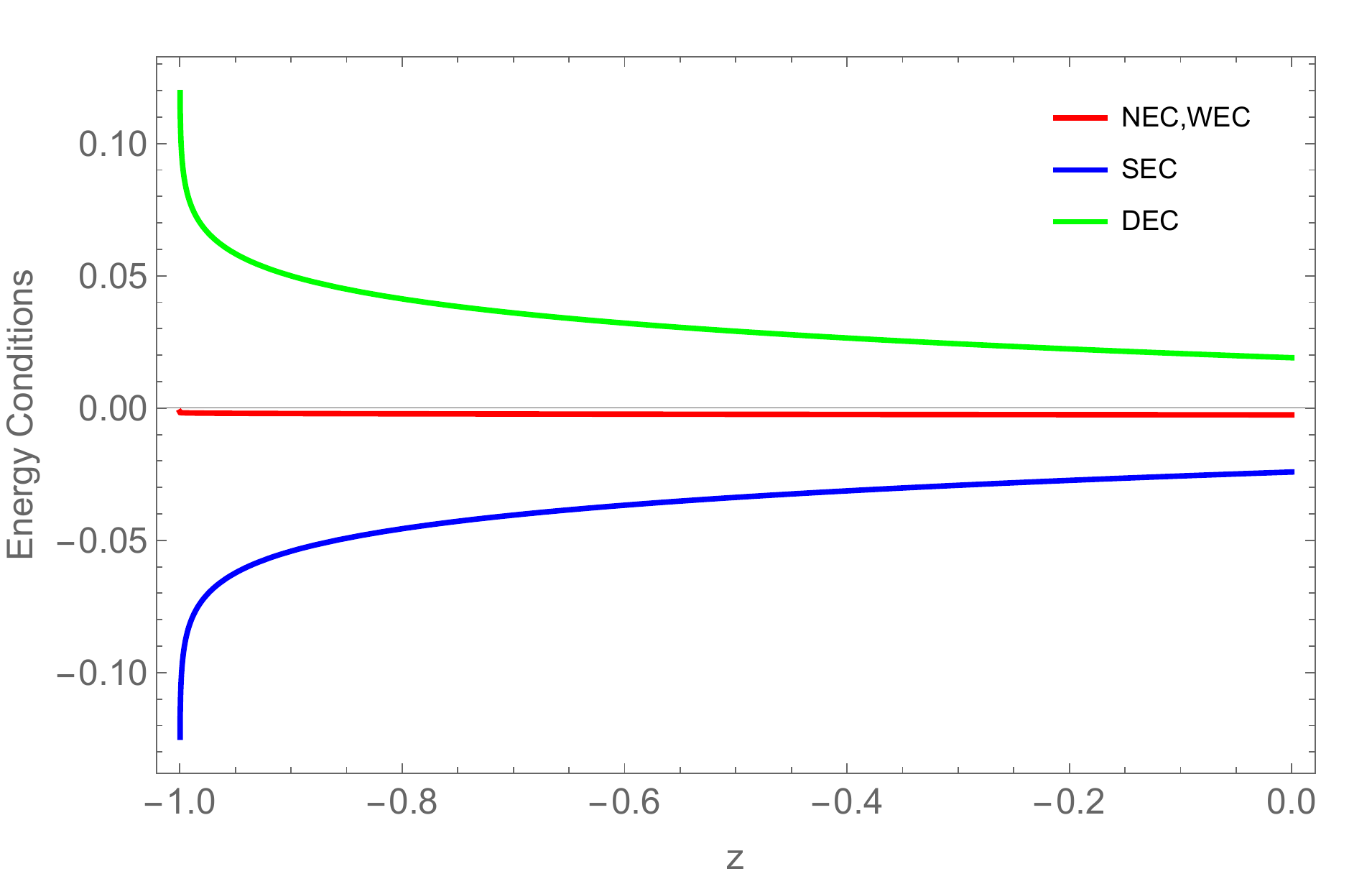}
\includegraphics[width=80mm]{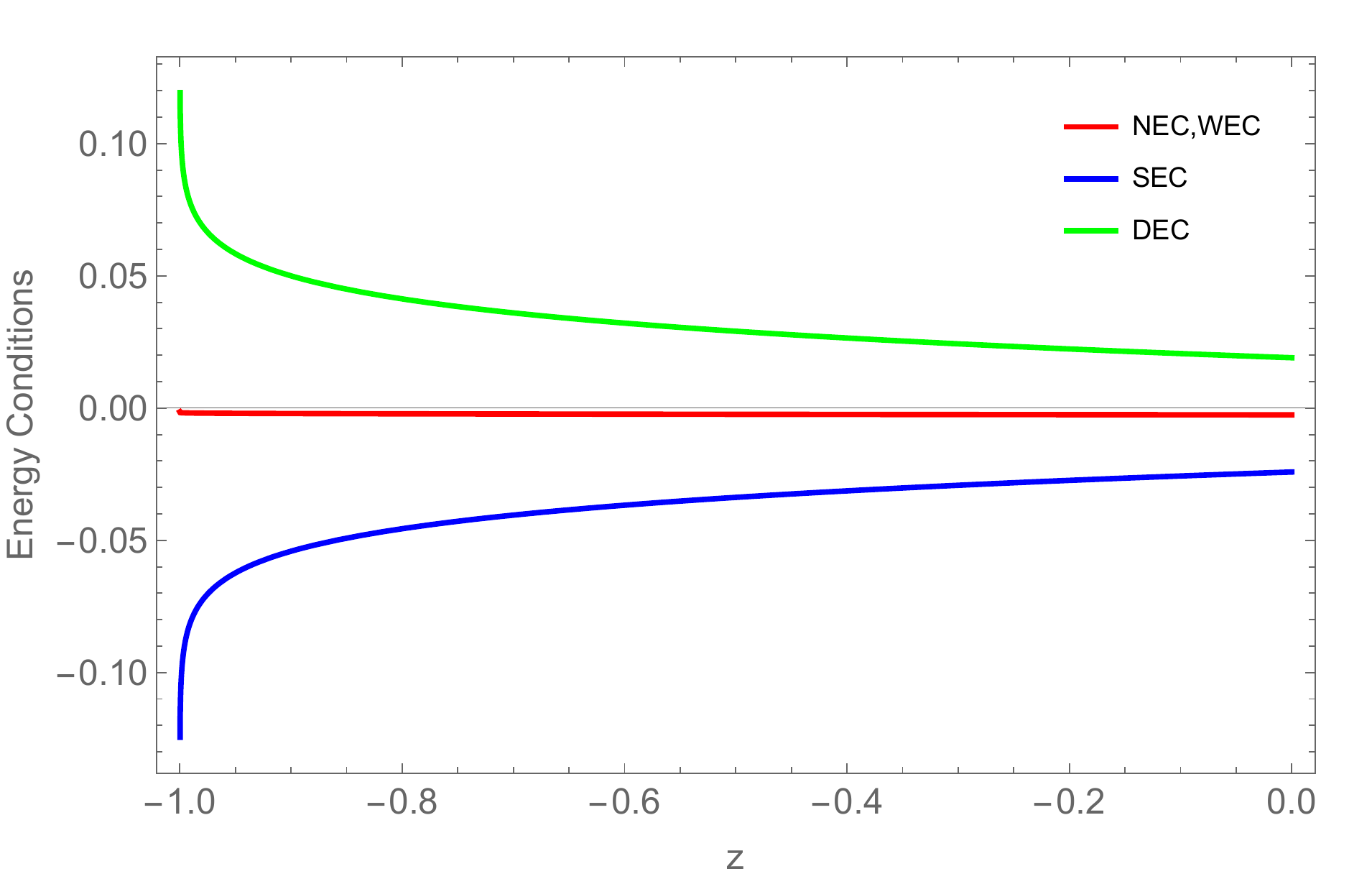}
\caption{Energy conditions vs redshift for varying $\alpha=0.042 $(left panel), $\alpha=0.044$ (right panel). The other parameter's values are, $A=0.4$, $\beta=-5.1$, $n=0.8$, $t_{\star}=10.1$., $\zeta=0.01$.} 
\label{Fig7}
\end{figure}

We can refer Eq. \eqref{eq:13a} to calculate the energy conditions of the exponential scale factor model as,
\begin{eqnarray}
\rho_{eff}+p_{eff}&=&-\frac{4 \alpha  \zeta }{t_{\star}^2}+\frac{\beta  t_{\star}^2 12^n (n-1) n \chi ^n \left(12 t_{\star}^2 \zeta  (n-2) t^2+t_{\star}^4-36 \zeta ^2 t^4\right)}{\left(t_{\star}^2+6 \zeta  t^2\right)^3}\nonumber \,, \\
\rho_{eff}+3p_{eff}&=&-\frac{12 \alpha  \zeta  \left(t_{\star}^2+2 \zeta  t^2\right)}{t_{\star}^4}+\frac{\beta  12^n (n-1) \chi ^n \left(-36 t_{\star}^2 \zeta ^2 (n+3) t^4+6 t_{\star}^4 \zeta  (2 n (3 n-5)-3) t^2+t_{\star}^6 (3 n-1)-216 \zeta ^3 t^6\right)}{\left(t_{\star}^2+6 \zeta  t^2\right)^3} \nonumber \,, \\
\rho_{eff}-p_{eff}&=&-\frac{\beta  12^n (n-1) \chi ^n \left(36 t_{\star}^2 \zeta ^2 (n-3) t^4+6 t_{\star}^4 \zeta  (2 (n-1) n-3) t^2+t_{\star}^6 (n-1)-216 \zeta ^3 t^6\right)}{\left(t_{\star}^2+6 \zeta  t^2\right)^3}+4 \alpha  \chi \,.
\end{eqnarray}
In FIG. ~\ref{Fig7}, we have presented the graphical behaviour of the energy conditions for two representative values of the model parameter $\alpha=0.042$ (left panel) and $\alpha=0.044$ (right panel). In both the cases, the DEC satisfied whereas SEC violates. The WEC violates  initially and at the late time it vanishes, which can be possible in the context of modified theories of gravity.

\subsection{Scalar Field Reconstruction}

The study of scalar field in particle physics encourage to study its applications in cosmology. The EoS parameter can be described as in section~\ref{sec:Scalar Field Reconstruction}, the equation of pressure and energy density is described in  Eqs.~\eqref{eq:14} and \eqref{eq:15} respectively. Cosmological solution for self-interacting potentials like simple power law potential, exponential potential, the simple logarithmic potential is studied and evolution equation has been calculated to study different state parameters and the generalized Lagrangian function for various potentials \cite{Chakrabarti17}. The approach of introducing a scalar field reconstruction technique in $f(T)$ gravity with the study of constant-roll scalar potential to obtain the Hubble evolution in teleparallel gravity has been studied in Ref.\cite{Awad18}. The expressions for squared slope of scalar field and  self interacting potentials can be written as,
\begin{eqnarray}
    \epsilon \dot{\phi}^2&=-\frac{4 \alpha  \zeta }{t_{\star}^2}+\frac{\beta  t_{\star}^2 12^n (n-1) n \chi ^n \left(12 t_{\star}^2 \zeta  (n-2) t^2+t_{\star}^4-36 \zeta ^2 t^4\right)}{\left(t_{\star}^2+6 \zeta  t^2\right)^3}\,,
\end{eqnarray}
\begin{eqnarray}
    2 V(\phi)&=-\frac{\beta  12^n (n-1) \chi ^n \left(36 t_{\star}^2 \zeta ^2 (n-3) t^4+6 t_{\star}^4 \zeta  (2 (n-1) n-3) t^2+t_{\star}^6 (n-1)-216 \zeta ^3 t^6\right)}{\left(t_{\star}^2+6 \zeta  t^2\right)^3}+4 \alpha  \chi\,.
\end{eqnarray}
\begin{figure}[!htp] 
\centering
\includegraphics[width=80mm]{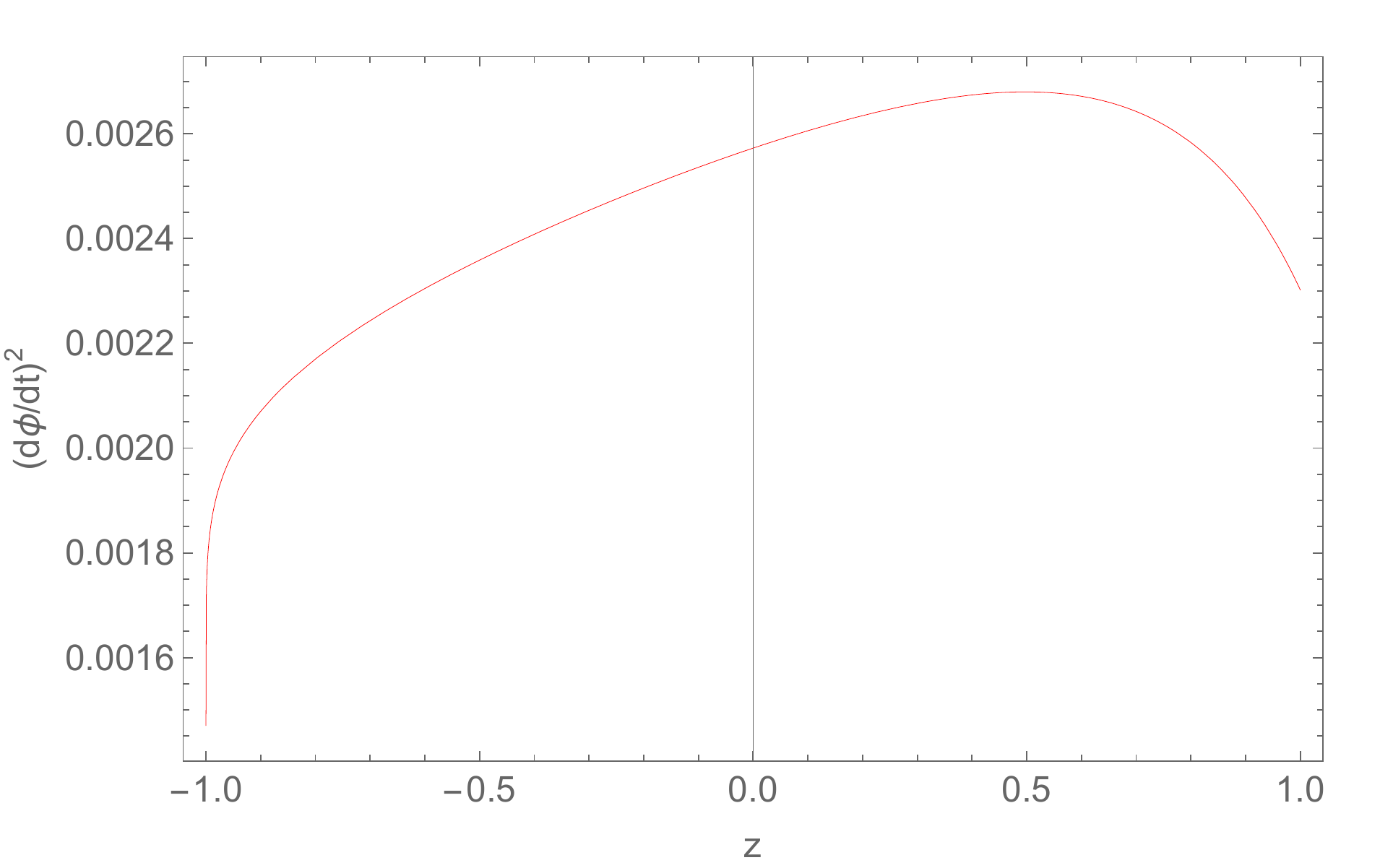}
\includegraphics[width=80mm]{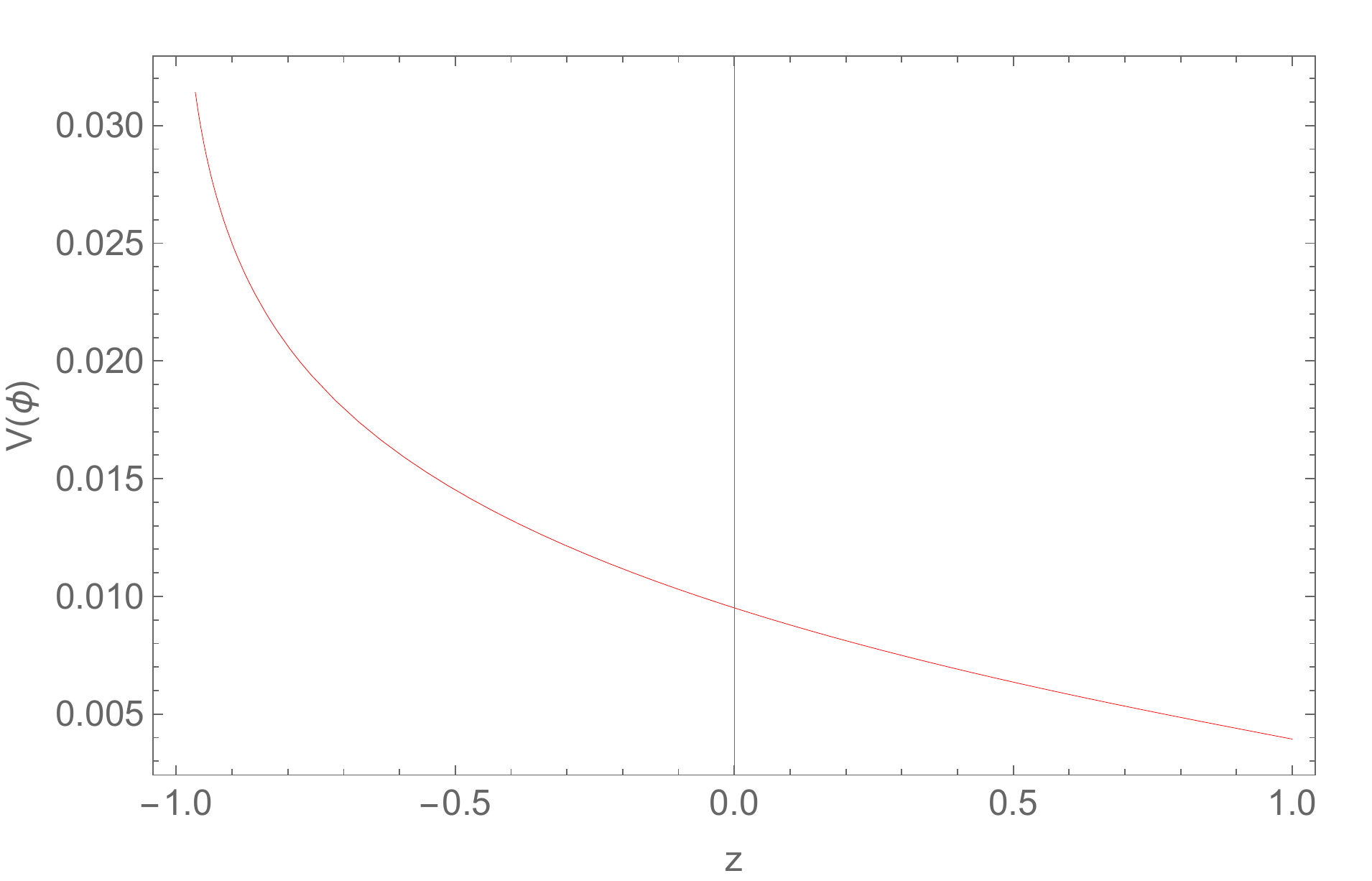}
\caption{Squared slope of reconstructed scalar field as a function of $\dot{\phi}^2$ (left panel) and $V(\phi)$ (right panel) vs redshift. The parameteric values are, $A=0.4$, $\alpha=0.042$,  $\beta=-5.1$, $n=0.8$, $t_{\star}=10.1$, $\zeta=0.01$.}
\label{Fig8}
\end{figure}
The squared slope remains in the positive domain and shows some increment at the initial time and then decreases, however around $z=0$, it gradually decreases [FIG. \ref{Fig8} (left panel)]. At the same time the potential function increases over time from a lower value [FIG. \ref{Fig8} (right panel)]. 

\subsection{Stability Analysis}

In this section, the stability analysis of exponential  scale factor based model has been  investigated. The formula for adiabatic speed of sound is discussed in section \ref{sec:Stability analysis}. The observational constraints on unified dark matter with constant speed of sound with CMB analysis of $\Lambda \alpha$CDM  model has been studied by Balbi et al \cite{Balbi07}. Mishra and Shaikh studied an observational parameters and stability analysis in extended teleparallel, $f(T)$ gravity \cite{Mishra20}. The adiabatic speed of sound for an exponential scale factor can be written as,
\begin{eqnarray}
    \frac{dp_{eff}}{d\rho_{eff}}&=\frac{4 \alpha  \zeta  \left(t_{\star}^2+6 \zeta  t^2\right)^4-\beta  t_{\star}^4 12^n (n-1) n \chi ^n \left(6 t_{\star}^4 \zeta  (2 n (2 n-7)+9) t^2+t_{\star}^6 (6 n-13)-108 t_{\star}^2 \zeta ^2 t^4-216 \zeta ^3 t^6\right)}{\frac{\beta  t_{\star}^8 12^n (n-1) n \chi ^{n+1} \left(12 t_{\star}^2 \zeta  (n-2) t^2+t_{\star}^4-36 \zeta ^2 t^4\right)}{\zeta }-4 \alpha  \zeta  \left(t_{\star}^2+6 \zeta  t^2\right)^4}\,.
\end{eqnarray}

\begin{figure}[!htp]
\centering
\includegraphics[width=90mm]{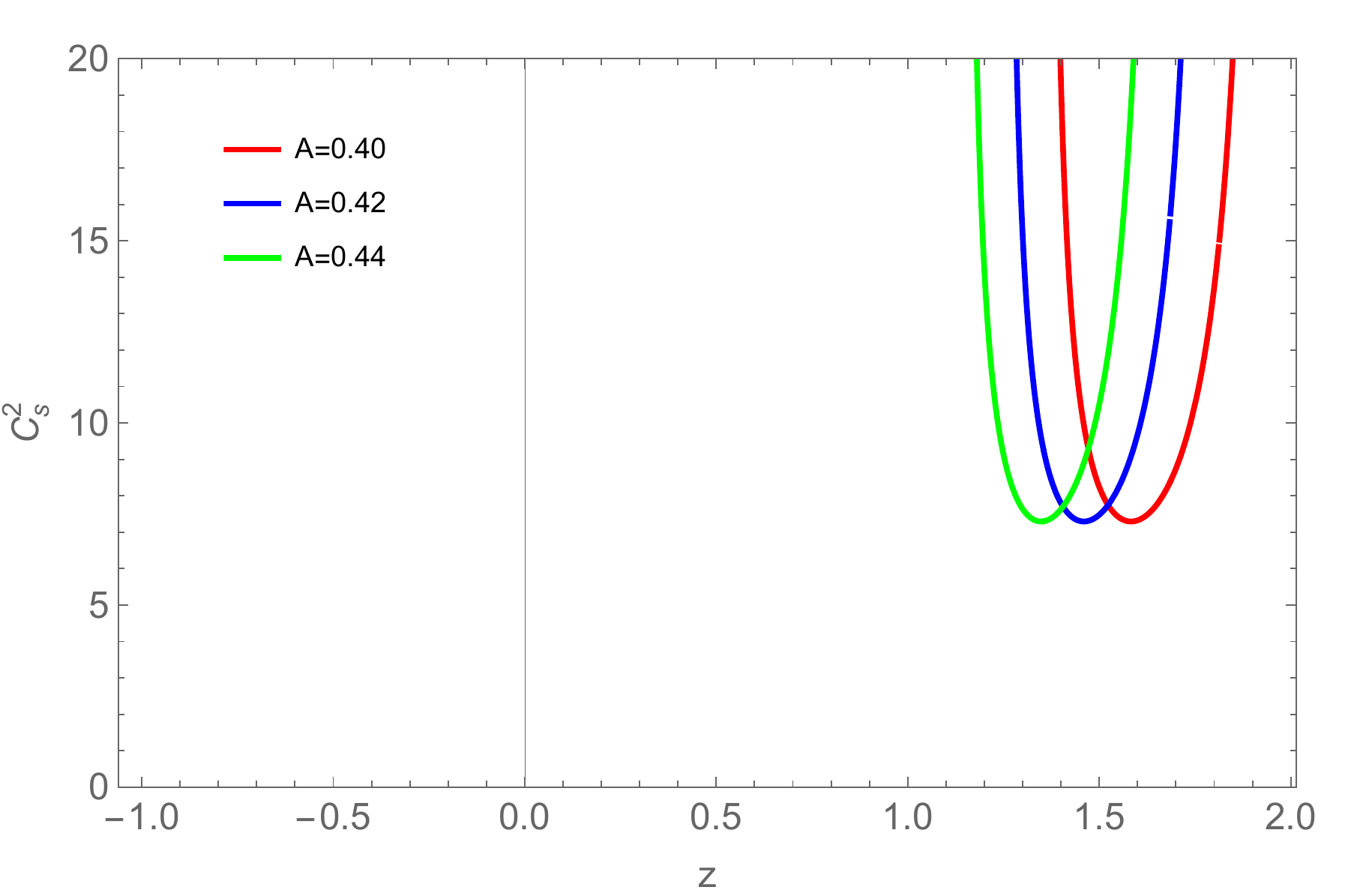}
\caption{Stability vs redshift with varying $A=0.40, 0.42, 0.44$. The other parameters value are, $\alpha=0.042$, $\beta=-5.1$, $n=0.8$, $t_{\star}=10.1$, $\zeta=0.01$.} 
\label{Fig9}
\end{figure}

The graph for adiabatic speed of sound in terms of redshift is plotted in FIG~\ref{Fig9}. The graph lies in the positive region, $C_s^2>0$ confirms the stable behaviour of the model.

\section{Conclusion}\label{sec:conclusion}

We have presented the cosmological models in  $f(T,B)$ gravity, a modified theory of gravity, where the Witzenb$\ddot{o}$ck connection has been used in stead of the usual Levi-Civita connection. A power law cosmology and exponential scale factor have been incorporated in the field equations of $f(T,B)$ gravity with the most general form of $\tilde{f}(T,B)=\alpha T+\beta B^n$. The effective pressure, effective energy density and effective EoS parameter are obtained at the background of homogeneous and isotropic space time. Both the models are showing late time accelerating behaviour. The EoS parameters of both models have been presented in a combination of appropriately chosen best fit values of the model and scale factor parameters. Though the evolution in each of the combination starts from different, but at late time all supports the $\Lambda$CDM behaviour. It has also been seen that the power law model remains in the $\Lambda$CDM phase in immediate future whereas in exponential case the evolution approaching to $\Lambda$CDM from the phantom phase. Though both the models are showing closer behaviour as compared to the observations, however the power law model obtained to be more promising. The details have been summarised in TABLE-I. The $f(T,B)$ gravity theory is capable to study the possible amplification of primordial magnetic fields\cite{Capozziello22} and also establish it's validity on the large scale by playing a vital role in the study of polarization of gravitational waves\cite{Capozziello20}. We wish to mention here that the hyperconical universe produces inhomogeneous metrics that are compatible with the observed expansion and approaches to the flat FLRW metric locally \cite{Monjo18}. It can be assumed as a local perturbation theory in inhomogeneous universes expanding to be consistent with the $\Lambda$CDM model regardless of the matter content. The model we have studied with the power law scale factor show concordance with the $\Lambda$CDM at the late time in the frame work of $f(T,B)$ gravity with flat FLRW space time. The result reported in this paper is compatible with the findings of the geometrical interpretation of the dark energy from projected hyperconical universe, however a detailed study may be taken up in future. \\

For the power law cosmology, the energy conditions are plotted for varying values of $\alpha$ and $\beta$. All the plots have shown the violation of strong energy conditions, which has been a prescription for the geometrical modified theories of gravity. In all the three plots of FIG~\ref{Fig3}, the DEC is satisfying whereas the WEC/NEC vanishes immediately after $z=0$. The violation of SEC further strengthen the validity of the model in the context of accelerating universe. The stability analysis enabled us to assess the generality of the assumptions made to frame the model. We have obtained that our model is showing stable behaviour (FIG~\ref{Fig5}) even if in different power of the scale factor function. The details has been summarised in TABLE-II. We reconstructed the scalar field in the context of modified gravity. The squared slope shows decreasing behaviour whereas the potential function shows positive behaviour and decreases gradually. We can conclude that the scalar field is model dependent.  \\

In the exponential scale factor case, the violation of SEC has been observed for the chosen value of the model parameters, whereas the DEC is satisfying. The NEC violates at initial time and vanishes at late time. The squared slope initially increases and after some time start decreasing and approaches to zero. At the same time, the potential function reduces in the positive domain.  The model shows the stability throughout the evolution. In conclusion, we can infer that the $f(T,B)$ gravity can be another extended gravity to investigate the late time cosmic acceleration issue. In addition, more involved research are required to investigate the other aspects of cosmology in this gravity. Some of the key results are listed in the following tables.
\begin{table}[htbp]
\caption{Comparison of $\omega_0$ value with the value of cosmological observations for both the models}
\centering
\begin{tabular}{ |c|c|c|c|c|c|c|  }
\hline
\multicolumn{3}{|c|}{Power Law}&\multicolumn{3}{|c|}{Exponential Law}& Observations\\
\cline{1-7}
Parameters & varying $h$ & $\omega_0$& Parameters & varying A & $\omega_0$& \multirow{ 4}{*}{\begin{tabular}{@{}c@{}}$\omega_0=-1.035 ^{+0.055}_{-0.059}$ \cite{Amanullah10}\\ \vspace{0.3cm}$\omega_0=-1.29 ^{+0.15}_{-0.12}$ \cite{Valentino16}\end{tabular}}\\
\cline{1-6}
$\alpha=0.01$, $\beta=-0.4$, $n=0.001$, $t_0=1.1$ & 0.4 & -0.9682  &{\begin{tabular}{@{}c@{}}$\beta=-5.1$, $n=0.8$, $t_{\star}=10.1$,\\ $\alpha=0.042,\zeta=0.01$\end{tabular}}&0.40 & -1.317 & \\
 & 0.5 & -0.9558 & & 0.42 &-1.339 &\\
  & 0.6 & -0.9520 & &0.44 &-1.350 &\\
\hline
\end{tabular}
\end{table}

\begin{table}[htbp]
\caption{Behaviour of energy conditions and stability behaviour}
\centering
\begin{tabular}{ |c|c|c|c|c| }
\hline
Test & \multicolumn{2}{|c|}{Power Law} & \multicolumn{2}{|c|}{Exponential Scale Factor}\\
\hline
Energy Conditions & Early Time ($z\gg 1$) & Late Time (z $\simeq$ -1) &Early Time ($z\gg 1$)& Late Time (z $\simeq$ -1)\\
\hline
DEC & Satisfied   & Satisfied  & Satisfied &Satisfied\\
WEC & Satisfied   & Vanishes & Violated & Vanishes\\
NEC & Satisfied   & Vanishes & Violated & Vanishes\\
SEC & Violated   & Violated & Violated & Violated\\
\hline
Stability & Stable   & Stable & Stable & Stable\\
\hline
\end{tabular}
\end{table}

\section*{Acknowledgement}

SAK acknowledges the financial support provided by University Grants Commission (UGC) through Senior Research Fellowship (UGC Ref. No.: 191620205335), to carry out the research work. BM acknowledges IUCAA, Pune, India for the academic support to carry out the research work. The authors are thankful to the honorable referee for the valuable comments and suggestions to improve the quality of paper.


\begin{thebibliography}{99}
\section*{References}
\bibitem{Perlmutter99} S. Perlmutter et al., ``Measurements of $\Omega$ and $\Lambda$ from 42 High-Redshift Supernovae," \href{https://dx.doi.org/10.1086/307221}{\textit{Astrophys. J.}, \textbf{517}, 565 (1999).} 

\bibitem{Riess98} A. G. Riess et al., ``Observational Evidence from Supernovae for an Accelerating Universe and a Cosmological Constant," \href{https://doi.org/10.1086/300499}{\textit{Astronomical J.}, \textbf{116}, 1009 (1998).}


\bibitem{Linder10} E. V. Linder, ``Einstein’s other gravity and the acceleration of the Universe," \href{http://dx.doi.org/10.1103/PhysRevD.81.127301}{\textit{Phys. Rev. D}, \textbf{81}, 127301 (2010).}

\bibitem{Weitzenbock23} R. Weitzenb\"{o}ck, ``Invariantentheorie. Von dr. Roland Weitzenböck," \href{http://name.umdl.umich.edu/ABV0733.0001.001}{ Noordhoff, Gronningen, (1923).}

\bibitem{Bahamonde21} S. Bahamonde et al.,``Teleparallel Gravity: From Theory to Cosmology,"\href{https://doi.org/10.1088/1361-6633/ac9cef}{ \textit{Rep. Pro. Phys.,} (Accepted for publication) (2022).}

\bibitem{Bajardi21} F. Bajardi, S. Capozziello, ``Noether symmetries and quantum cosmology in extended teleparallel gravity," \href{https://doi.org/10.1142/S0219887821400028}{ \textit{Int. J. Geo. Meth. Mod. Phys.}, \textbf{18}, 2140002 (2021).}

\bibitem{Basilakos13} S. Basilakos et al.,``Noether symmetries and analytical solutions in $f(T)$ cosmology: A complete study,"\href{http://dx.doi.org/10.1103/PhysRevD.88.103526}{ \textit{Phys. Rev. D}, \textbf{88}, 103526 (2013).}

\bibitem{Myrzakulov11} R. Myrzakulov, ``Accelerating universe from $f(T)$ gravity," \href{https://doi.org/10.1140/epjc/s10052-011-1752-9}{\textit{Eur. Phys. J. C}, \textbf{71}, 1 (2011).} 

\bibitem{Capozziello11} S. Capozziello et al.,``Cosmography in $f(T)$ gravity," \href{http://dx.doi.org/10.1103/PhysRevD.84.043527} {\textit{Phys. Rev. D}, \textbf{84}, 043527 (2011).}

\bibitem{Briffa20}  R. Briffa et al., ``Constraining teleparallel gravity through Gaussian processes," \href{https://doi.org/10.1088/1361-6382/abd4f5}{ \textit{Class. Quant. Grav.}, \textbf{38}, 055007 (2021).} 

\bibitem{Said21} J. Levi Said et al.,``Reconstructing teleparallel gravity with cosmic structure growth and expansion rate data," \href{https://doi.org/10.1088/1475-7516/2021/06/015}{\textit{JCAP}, \textbf{06}, {015} (2021).} 

\bibitem{Cai20} Y. F. Cai, M. Khurshudyan, E. N. Saridakis, ``Model-independent Reconstruction of $f(T)$ Gravity from Gaussian Processes," \href{https://doi.org/10.3847/1538-4357/ab5a7f}{\textit{Astrophys. J.,} \textbf{888}, {62} (2020).}

\bibitem{Cai16} Y. F. Cai et al., ``$f(T)$ teleparallel gravity and cosmology," \href{http://dx.doi.org/10.1088/0034-4885/79/10/106901}{ \textit{Rept. Prog. Phys.}, \textbf{79}, {106901} (2016).}

\bibitem{Awad17} A. Awad, G. Nashed, ``Generalized teleparallel cosmology and initial singularity crossing," \href{http://dx.doi.org/10.1088/1475-7516/2017/02/046}{ \textit{JCAP}, \textbf{02}, {046} (2017).} 

\bibitem{Channuie18} P. Channuie, D. Momeni, ``Noether symmetry in a nonlocal $f(T)$ gravity,"\href{https://doi.org/10.1016/j.nuclphysb.2018.08.016}{\textit{ Nucl. Phys. B}, \textbf{935}, {256} (2018).}

\bibitem{Mirza19} B. Mirza, F. Oboudiat, ``Mimetic $f(T)$ teleparallel gravity and cosmology," \href{https://doi.org/10.1007/s10714-019-2576-4}{ \textit{Gen. Rel. Grav.} \textbf{51}, {96} (2019).} 

\bibitem{Golovnev20} A. Golovnev, M. J. Guzman, ``Bianchi identities in $f(T)$ gravity: Paving the way to confrontation with astrophysics," \href{https://doi.org/10.1016/j.physletb.2020.135806}{ \textit{Phys. Lett. B}, \textbf{810}, {135806} (2020).}

\bibitem{Ferraro08} R. Ferraro, F. Fiorini, ``Born-Infeld gravity in Weitzenb\"{o}ck spacetime,"\href{http://dx.doi.org/10.1103/PhysRevD.78.124019}{\textit{ Phys. Rev. D}, \textbf{78}, {124019} (2008).}

\bibitem{Li11} B. Li, T. P. Sotiriou, J. D. Barrow, ``$f(T)$
  gravity and local Lorentz invariance,"\href{http://dx.doi.org/10.1103/PhysRevD.83.064035}{\textit{Phys. Rev. D}, \textbf{83}, {064035} (2011).} 

\bibitem{Ferraro11} R. Ferraro, F. Fiorini, ``Spherically symmetric static spacetimes in vacuum 
$f(T)$ gravity,"\href{http://dx.doi.org/10.1103/PhysRevD.84.083518}{\textit{ Phys. Rev. D}, \textbf{84}, {083518} (2011).}

\bibitem{Nashed13} G. G. L. Nashed, ``A special exact spherically symmetric solution in $f(T)$ gravity theories," \href{https://doi.org/10.1007/s10714-013-1566-1}{\textit{Gen. Rel. Grav.}, \textbf{45}, {1887} (2013).}

\bibitem{Izumi13} K. Izumi, Y. C. Ong, ``Cosmological perturbation in $f(T)$ gravity revisited," \href{https://doi.org/10.1088/1475-7516/2013/06/029}{ \textit{JCAP} \textbf{06}, {029} (2013).}

\bibitem{Paliathanasis14} A. Paliathanasis et al., ``New Schwarzschild-like solutions in 
$f(T)$
 gravity through Noether symmetries,"\href{https://doi.org/10.1103/PhysRevD.89.104042}{ \textit{Phys. Rev. D}, \textbf{89}, {104042} (2014).}

\bibitem{Bejarano17} C. Bejarano, R. Ferraro, M. J. Guzman, ``McVittie solution in $f(T)$ gravity," \href{https://doi.org/10.1140/epjc/s10052-017-5394-4}{ \textit{Eur. Phys. J. C}, \textbf{77}, {825} (2017).}

\bibitem{Krssak16} M. Krssak, E. N. Saridakis, ``The covariant formulation of $f(T)$ gravity," \href{https://doi.org/10.1088/0264-9381/33/11/115009}{ \textit{Class. Quant. Grav.}, \textbf{33}, {115009} (2016).}

\bibitem{Bose20} A. Bose, S. Chakraborty, ``Cosmic evolution in $f(T)$ gravity theory," \href{https://doi.org/10.1142/S021773232050296X}{ \textit{Mod. Phys. Lett. A}, \textbf{35}, {2050296} (2020).}

\bibitem{Jimenez21}  J. B. Jimenez et al., ``Minkowski space in 
$f(T)$
 gravity," \href{https://doi.org/10.1103/PhysRevD.103.024054}{ \textit{Phys. Rev. D}, \textbf{103}, {024054} (2021).}

\bibitem{Duchaniya22} L. K. Duchaniya, S. V. Lohakare, B. Mishra, S. K. Tripathy, ``Dynamical stability analysis of accelerating $f(T)$ gravity models," \href{https://doi.org/10.1140/epjc/s10052-022-10406-w}{\textit{Eur. Phys. J. C}, \textbf{82}, 448 (2022).}

\bibitem{Capozziello15} S. Capozziello, O. Luongo, E. N. Saridakis, ``Transition redshift in 
$f(T)$
 cosmology and observational constraints," \href{http://dx.doi.org/10.1103/PhysRevD.91.124037}{ \textit{Phys. Rev. D}, \textbf{91}, 124037 (2015).}

\bibitem{Capozziello16} S. Capozziello, M. D. Laurentis, K. F. Dialektopoulos, ``Noether symmetries in Gauss–Bonnet-teleparallel cosmology," \href{https://doi.org/10.1140/epjc/s10052-016-4491-0}{ \textit{Eur. Phys. J. C}, \textbf{76}, 629 (2016).}

\bibitem{Kadam22EPJC} S. A. Kadam, B. Mishra,  J. L. Said, ``Teleparallel scalar-tensor gravity through cosmological dynamical systems," \href{https://doi.org/10.1140/epjc/s10052-022-10648-8}{\textit{Eur. Phys. J. C}, \textbf{82}, 680 (2022).}


\bibitem{Buchdahl70} H. A. Buchdahl,  ``Non-Linear Lagrangians and Cosmological Theory," \href{https://doi.org/10.1093/mnras/150.1.1}{\textit{Mon. Not. Roy. Astron. Soc.}, \textbf{150}, {1} (1970).}

\bibitem{Bahamonde15} S. Bahamonde, C.G. B$\ddot{o}$hmer, M. Wright, ``Modified teleparallel theories of gravity."  \href{	https://doi.org/10.1103/PhysRevD.92.104042}{\textit{Phys. Rev. D}, \textbf{92}, 104042 (2015).}

\bibitem{Bahamonde17}  S. Bahamonde, S. Capozziello, ``Noether symmetry approach in $f(T, B)$ teleparallel cosmology." \href{https://doi.org/10.1140/epjc/s10052-017-4677-0}{\textit{Eur. Phys. J. C}, \textbf{77}, {107} (2017).}

\bibitem{Capozziello22} S. Capozziello, A. Carleo, G. Lambiase, ``The amplification of cosmological magnetic fields in extended $f(T,B)$ teleparallel gravity," \href{https://doi.org/10.1088/1475-7516/2022/10/020}{\textit{JCAP}, \textbf{2022}, {020} (2022).}

\bibitem{Capozziello20} S. Capozziello, M. Capriolo, L. Caso, ``Weak field limit and gravitational waves in $f(T, B)$ teleparallel gravity," \href{https://doi.org/10.1140/epjc/s10052-020-7737-9}{\textit{Eur. Phys. J. C}, \textbf{80}, {156} (2020).}

\bibitem{Bahamonde18} S. Bahamonde, M.Zubair, G. Abbas, ``Thermodynamics and cosmological reconstruction in $f(T,B)$ gravity," \href{https://doi.org/10.1016/j.dark.2017.12.005} {\textit{Phys. Dark Uni.}, \textbf{19}, 78 (2018).}

\bibitem{Caruana20} M. Caruana, G. Farrugia, J. L. Said, ``Cosmological bouncing solutions in $f(T, B)$ gravity," \href{https://doi.org/10.1140/epjc/s10052-020-8204-3} {\textit{Eur. Phys. J. C}, \textbf{80}, 640 (2020).} 

\bibitem{Franco20} G. A. R. Franco, C. E. Rivera, J. L. Said, ``Stability analysis for cosmological models in $f(T, B)$ gravity," \href{https://doi.org/10.1140/epjc/s10052-020-8253-7}{ \textit{Eur. Phys. J. C}, \textbf{80}, {677} (2020).}

\bibitem{Rivera20} C. E. Rivera, J. L. Said, ``Cosmological viable models in $f(T, B)$ theory as solutions to the $H_0$ tension,"\href{https://doi.org/10.1088/1361-6382/ab939c} {\textit{Class. Quan. Grav.}, \textbf{37}, 165002 (2020).}

\bibitem{Pourbagher20} A. Pourbagher, A. Amani, ``Thermodynamics of the viscous $f(T,B)$ gravity in the new agegraphic dark energy model," \href{https://doi.org/10.1142/S0217732320501667}{ \textit{Mod. Phys. Lett. A}, \textbf{35}, {2050166} (2020).}

\bibitem{Zubair20} M. Zubair, L. R. Durrani, ``A study of the cosmologically reconstructed $f(T, B)$ gravity from the cosmological jerk parameter," \href{https://doi.org/10.1140/epjp/s13360-020-00681-5}{ \textit{Eur. Phys. J. Plus}, \textbf{135}, {668} (2020).}

\bibitem{Moreira21} A. R. P. Moreira et al., ``Thick brane in 
$f(T,B)$
 gravity," \href{https://doi.org/10.1103/PhysRevD.103.064046}{ \textit{Phys. Rev. D}, \textbf{103}, {064046} (2021).}

\bibitem{Sahlu21}  S. Sahlu et al., ``Inflationary constraints in teleparallel gravity theory," \href{https://doi.org/10.1142/S0219887821500274}{ \textit{Int. J. Geom. Meth. Mod. Phys.}, \textbf{18}, {2150027} (2021).}

\bibitem{Bhattacharjee21} S. Bhattacharjee, ``Constraining $f(T, B)$ teleparallel gravity from energy conditions," \href{https://doi.org/10.1016/j.newast.2020.101495}{ \textit{New Astron.}, \textbf{83}, {101495} (2021).}

\bibitem{Kadam22} S. A. Kadam, B.Mishra, S.K. Tripathy, ``Dynamical features of $f(T,B)$ cosmology," \href{https://doi.org/10.48550/arXiv.2206.00430}{\textit{Mod. Phys. Lett. A}, \textbf{37}, {17} (2022).}

\bibitem{Clifton12} T. Clifton et al., ``Modified gravity and cosmology," \href{https://doi.org/10.1016/j.physrep.2012.01.001}{ \textit{Phys. Rept.}, \textbf{513}, {1} (2012).}

\bibitem{Linde82}  A. D. Linde, ``A new inflationary universe scenario: A possible solution of the horizon, flatness, homogeneity, isotropy and primordial monopole problems," \href{https://doi.org/10.1016/0370-2693(82)91219-9}{\textit{Phys. Lett. B}, \textbf{108}, {389} (1982).}

\bibitem{Guth81} A. H. Guth, ``Inflationary universe: A possible solution to the horizon and flatness problems," \href{https://doi.org/10.1103/PhysRevD.23.347}{\textit{Phys. Rev. D}, \textbf{23}, {347} (1981).}

\bibitem{Hawking73} S. W. Hawking, G. F. R. Ellis, ``The Large Scale Structure of Space-Time," \href{https://ui.adsabs.harvard.edu/abs/1975lsss.book.....H}{Cambridge university press, \textbf{1}, (1975).}

\bibitem{Matt97} M. Visser, ``Energy Conditions in the Epoch of Galaxy Formation," \href{https://ui.adsabs.harvard.edu/link_gateway/1997Sci...276...88V/doi:10.1126/science.276.5309.88}{\textit{Science (New York, NY)} \textbf{276}, {88}, (1997).}


\bibitem{Charters01} T. C. Charters, A. Nunes, J. P. Mimoso, ``Stability analysis of cosmological models through Lyapunov's method," \href{https://ui.adsabs.harvard.edu/link_gateway/2001CQGra..18.1703C/doi:10.1088/0264-9381/18/9/307}{\textit{Class. Quant. Grav.}, \textbf{18}, {1703} (2001).}

\bibitem{Balbi07} A. Balbi, M. Bruni, C. Quercellini, ``$\Lambda\alpha$DM
 : Observational constraints on unified dark matter with constant speed of sound," \href{https://link.aps.org/doi/10.1103/PhysRevD.76.103519}{\textit{Phys. Rev. D}, \textbf{76}, {103519} (2007).}

\bibitem{Xu13} L. Xu, ``Unified dark fluid with constant adiabatic sound speed: Including entropic perturbations," \href{https://ui.adsabs.harvard.edu/link_gateway/2013PhRvD..87d3503X/doi:10.1103/PhysRevD.87.043503}{ \textit{Phys. Rev. D}, \textbf{87}, {043503} (2013).}

\bibitem{Sharif17} M. Sharif, A. Ikram, ``Stability analysis of some reconstructed cosmological models in $f(G,T)$ gravity," \href{https://doi.org/10.1016/j.dark.2017.05.001}{ \textit{Phys. Dark Univ.}, \textbf{17}, {1} (2017).}
 
\bibitem{Shah19} P. Shah, G. C. Samanta, ``Stability analysis for cosmological models in $f(R)$ gravity using dynamical system analysis," \href{https://doi.org/10.1140/epjc/s10052-019-6934-x} {\textit{Eur. Phys. J. C}, \textbf{79}, {414} (2019).}

\bibitem{Mishra21} B. Mishra et al., ``Stability analysis of two-fluid dark energy models," \href{https://doi.org/10.1088/1402-4896/abdf82}{ \textit{Phys. Scripta}, \textbf{96}, {045006} (2021).}

\bibitem{Chakrabarti17} S. Chakrabarti, J. L. Said, G. Farrugia, ``Some aspects of reconstruction using a scalar field in $f(T)$ gravity," \href{https://doi.org/10.1140/epjc/s10052-017-5404-6}{\textit{Eur. Phys. J. C}, \textbf{77}, {815},(2017).}

\bibitem{Awad18} A. Awad et al., ``Constant-roll inflation in $f(T)$ teleparallel gravity," \href{https://doi.org/10.1088/1475-7516/2018/07/026}{\textit{JCAP}, \textbf{2018}, {026} (2018).}

\bibitem{Mishra20} A. Y. Shaikh, B. Mishra, ``Analysis of observational parameters and stability in extended teleparallel gravity," \href{https://doi.org/10.1142/S0219887820501583}{\textit{Int. J. Geom. Meth. Mod. Phys.,} \textbf{17}, 2050158 (2020).} 

\bibitem{Monjo18} R. Monjo, ``Geometric interpretation of the dark energy from projected hyperconical universes," \href{https://link.aps.org/doi/10.1103/PhysRevD.98.043508}{\textit{Phys. Rev. D}, \textbf{98}, 043508 (2018).}

\bibitem{Amanullah10} R. Amanullah et al., ``Spectra and Hubble space telescope light curves of six type Ia supernovae at 0.511 $< $z $< $1.12 and the union2 compilation*," \href{https://doi.org/10.1088/0004-637x/716/1/712}{ \textit{Astrophys. J.}, \textbf{716}, 712 (2010).}

\bibitem{Valentino16} E.D. Valentino, A. Melchiorri, J. Silk, ``Reconciling Planck with the local value of $H_0$ in extended parameter space," \href{https://doi.org/10.1016/j.physletb.2016.08.043}{\textit{Phys. Lett. B}, \textbf{761}, 242 (2016).}



\end{thebibliography}
\end{document}